\documentclass[useAMS,usenatbib,usegraphicx]{mn2e}

\usepackage{amssymb}        
\usepackage{amsmath}        

\title{H$\alpha$ Star Formation Rates in Massive Galaxies at $z \sim 1$}
\author[Twite et al.]{Jonathan W. Twite$^{1}$\thanks{E-mail:ppxjwt@nottingham.ac.uk}, Christopher J. Conselice$^{1}$, Fernando Buitrago$^{1}$,
\newauthor Kai Noeske$^{3}$, Benjamin J. Weiner$^{2}$, Jose A. Acosta-Pulido$^{4,5}$, Amanda E. Bauer$^{1}$ \\
$^{1}$University of Nottingham, School of Physics \& Astronomy, Nottingham, NG7 2RD UK \\
$^{2}$Steward Observatory, University of Arizona, Tucson, AZ 85721, USA \\
$^{3}$Harvard-Smithsonian Center for Astrophysics, 60 Garden Street, Cambridge, MA 02138, USA \\
$^{4}$Instituto de Astrof\'{\i}sica de Canarias (IAC), V\'{\i}a L\'{a}ctea s/n, La Laguna, E-38205, Spain \\
$^{5}$Departamento de Astrof\'{\i}sica, Facultad de F\'{\i}sica, Universidad de la Laguna, Astrof\'{\i}sico Fco. S\'{a}nchez s/n, La Laguna, E-38207, Spain}

 \def\msol{M$_{\odot}\,$}
 \def\GALEX{\textit{GALEX}\,}
 \def\Spitzer{\textit{Spitzer}\,}
 \def\HST{\textit{Hubble Space Telescope}\,}
 \def\WHT{\textit{William Herschel Telescope}}
 \def\LIRIS{\textit{LIRIS}~}

\begin{document}

  \date{Accepted ; Received ; in original form}
  \pagerange{\pageref{firstpage}--\pageref{lastpage}} \pubyear{2002}

  \maketitle

  \label{firstpage}

  \begin{abstract}

    We present a near-infrared spectroscopic study of a stellar mass selected sample of galaxies at $z \sim 1$ utilising the LIRIS multi-object spectrograph on the William Herschel Telescope.  We detect continuum, and the H$\alpha$ line for our sample, which is one of the better direct tracers of star formation in external galaxies.  We spectroscopically measure the H$\alpha$ emission from 41 massive (M$_{*}>10^{10.5}$\,\msol) galaxies taken from the POWIR Survey with spectroscopic redshifts 0.4 $<$ z$_{\rm{spec}}$ $<$ 1.4.  We correct our H$\alpha$ fluxes for dust extinction by using multi-wavelength data,  and investigate SFR trends with mass and colour.   We find a drop in the fraction of massive galaxies with M$_{*} > 10^{11}$\,\msol which are detected in H$\alpha$ emission at $z <0.9$.   We furthermore find that the fraction of galaxies with H$\alpha$ emission drops steadily and significantly with redder $(U-B)$ colours at z $\sim 1$, and that the SSFR drops with increasing $(U-B)$ colour for galaxies at all masses.   By investigating the SFR-mass relation we find that the SFR is roughly constant with mass, in possible contrast to previous work, and that the specific star formation rate (SSFR) is lower in the most massive galaxies.   The scatter in the SFR vs. mass relationship is very small for those systems with ongoing star formation which suggests that star formation in the most massive galaxies at $z \sim 1$ shuts off rather abruptly over $< 1$\,Gyr, without an obvious gradual decline.  We furthermore investigate the SFR as a function of $(U-B)$ colour divided into different mass bins, revealing a tracer of the epoch of transition from star forming to passive, as a form of star formation ``downsizing''.  This suggests that the shut off of star formation occurs before the change in a galaxy's colour.  We find that galaxy stellar mass is the primary driving mechanisms behind the star formation history for these galaxies and discuss several possible mechanisms for regulating this process.


  \end{abstract}
    
  \begin{keywords}
    galaxies: star formation - galaxies: evolution - galaxies: formation
  \end{keywords}

  \section{Introduction}
    
    
    It has been long known that galaxies have evolved over the history of the universe, and a major branch of recent astronomy has been to quantify how the galaxies we see in the present day came to be as they are now.  It is impossible to observe how a given galaxy evolves over cosmologically-significant time scales, but its evolution can be inferred either on a galaxy by galaxy basis from the present state of individual galaxies -- for example the star formation history (e.g. Maraston et al. 2006; Noeske et al. 2007b), or the merger history (e.g. Conselice 2003; Conselice et al. 2009); or from the changes to the properties of a population of galaxies at different times.
    
    Massive galaxies are a major component of a number of branches of astronomy, being a pivotal test for, amongst others, dark matter theories and galaxy formation models.  Multiple constraints have now been put on when the most massive (M$_{*}>10^{10}$\,\msol) galaxies formed (e.g., Fontana et al. 2004; Glazebrook et al. 2004; Bundy et al. 2006; Conselice et al. 2007, 2011) and it is seen that $z \sim1$ is an era of transition when massive galaxies shut off their star formation; their mass for the most part, already in place (Glazebrook et al. 2004; Bundy et al. 2006; Conselice et al. 2007; P\'{e}rez-Gonz\'{a}lez et al. 2008a; Ilbert et al. 2010).  The star formation in these most massive galaxies was significantly higher before this epoch, and there appears to be evidence of star formation rates well in excess of those found in today's universe (e.g. Hopkins 2004; P\'{e}rez-Gonz\'{a}lez et al. 2008b).
    
    Within this transition epoch, there is increasing evidence that the shut off in star formation is not triggered simultaneously across the universe, but is highly dependent on the masses of the galaxies.  In a process that has been named `downsizing', it has been suggested that the most massive galaxies finish their star formation earlier than less massive systems, a progression observed in broad-band colours (e.g. Bundy et al. 2006), radio observations (Hopkins 2004), infrared observations (e.g. Papovich et al. 2006; Bell et al. 2007) and emission line surveys (e.g. Heavens et al. 2004; Panter et al. 2007).
    
    Measurements of high-redshift star formation rates (hereafter SFRs) have until recently, been dogged by many problems.  The more favoured emission line diagnostic, the H$\alpha$ line is only available via optical spectroscopy at $z \lesssim$ 0.4 and so the more troublesome [OII] line has been regularly used at higher redshifts (Gallagher, Hunter \& Bushouse  1989; Kennicutt 1998a; Rosa-Gonz\'{a}lez, Terlevich \& Terlevich 2002; Charlot et al. 2002; Kewley, Geller \& Jansen 2004).  The [OII] line is affected more by metalicity and due to its bluer wavelength, is more influenced by dust attenuation than the H$\alpha$ line (Jansen, Franx \& Fabricant 2001; Charlot et al. 2002).  As a result, it is never entirely certain whether high-$z$ SFRs derived from [OII] are completely compatible with low-$z$ SFRs derived from other more accurate methods.
    
    Infrared (IR) SFR tracers, for example 24\,$\mu$m emission (Lonsdale Persson \& Helou 1987; Rowan-Robinson \& Crawford 1989; Kennicutt 1998a; Calzetti et al. 2007) is also commonly used as a `proxy' SFR diagnostic.  Emission from SFR regions is absorbed by dust and re-radiated as IR radiation, allowing a conversion to be calibrated to retrieve the original SFR.  This method has however, problems in measuring star formation rates. For example, using current technology, spectral line tracers are also able to measure SFRs to an order of magnitude lower than the \Spitzer MIPS IR instrument, and are more sensitive at higher redshifts (see for example Noeske et al. 2007a).  Other problems include the fact that the extrapolation from a few sub-mm and radio luminosities to a bolometric IR luminosity is subject to many uncertainties, especially at high redshift (Smail, Ivison \& Blain 1997; Chapman et al. 2005) and systematic changes (for example Papovich et al. 2007); contamination by an active galactic nuclei (AGN) (see Daddi et al. 2007b); and possibly cirrus emission comming from older stars (Lonsdale Persson \& Helou 1987; Helou 1986; Kennicutt 1998a).  Radio SFRs, another SFR measure, often relies on stacking of data sources which may lead to complications in interpretting results (e.g. Ivison et al. 2010).
    
    Each measure of star formation tracks a different element of the stellar evolution cycle, each with its own precise definitions (See Kennicutt 1998a for a review of what different wavelengths actually measure) and therefore unique biases.  It is important therefore to get a measure of whether the apparent changes in global SFR are actually due to a change in the SFR of galaxies, or because the different measurements at different redshifts are measuring different processes.  The H$\alpha$ line is the best tracer of this due to the relatively low attenuation it suffers from dust compared to other emission line tracers, the lack of any strong dependency on metalicity, and because its emission is associated largely with only star forming regions.  It is also a direct measurement of star formation, and measures instantaneous star formation during the last 20\,Myr within the galaxy.  Attenuation is still a significant problem however -- it is variable galaxy-to-galaxy due to its dependence on the amount of dust in a galaxy, and can effect the emission by a factor of three or more.
    
    At redshift $z \gtrsim$ 0.4, the H$\alpha$ emission line shifts into the near-Infrared (NIR) and different technologies are required to observe it.  In the last decade there have been a number of IR instruments built for telescopes and with these new instruments, we can now begin to measure this emission to higher redshifts than before.  There have been a few small investigations into H$\alpha$ emission at $z \sim1$, however most have been hampered by a very high detection limit, or the low number of galaxies studied.  Glazebrook et al (1999) presented the first NIR spectroscopy to detect H$\alpha$ emission from galaxies at $z$ = 1.  They studied 13 field galaxies drawn from the Canada-France Redshift Survey and compare their derived SFRs with observations of the rest-frame UV emission.  They find that the H$\alpha$ SFR density measurements are roughly three times larger than the ones derived using the Madau, Pozzetti \& Dickinson (1996) UV relation.  Doherty et al (2006) followed from works by Yan et al. (1999), Hopkins, Connolly \& Salay (2000) and Tresse et al. (2002) examining the SFR density at $z \sim1$.  Doherty et al. took spectra of 38 galaxies within the \textit{Hubble Deep Field North} and stacked their spectra in order to measure the total star formation and hence the SFR density.  They estimate a lower limit for the total SFR in their stacked sample to be $\sim$ 312\msol yr$^{-1}$ but do not measure individual SFRs.  Both Doherty et al. and Yan et al. derive dust corrections by comparison with UV measurements.  Tresse et al. (2002) used a mixture of Balmer ratios (H$\alpha$/H$\beta$ and H$\gamma$/H$\beta$) where available to measure a dust correction, in other galaxies they used a fixed attenuation.  Most recently, Rodr\'{\i}guez-Eugenio et al. (2007) measured 16 star forming galaxies at redshift 0.8 $< z <$ 1.0 from the DEEP2 survey with the \LIRIS camera on the \WHT, but had problems removing the sky lines from their data.  Garn et al. (2010) used narrow band imaging to measure H$\alpha$ emission from 477 sources in the HiZELS survey with a SFR detection limit of 2.15\,\msol\,yr$^{-1}$.  They used other available multiwavelength imaging to calculate the dust attenuation using a number of methods.  Sobral et al. (2010) further investigate the SFR-mass relation and specific SFR-mass relation of these galaxies, however focusing on environmental effects.
    
    None of these previously mentioned surveys have however used H$\alpha$ spectroscopy to investigate trends and relations found with other SFR tracers, nor do they use a mass selected sample.  We present the first such study using the \LIRIS spectrograph on the \WHT\ to obtain near-Infrared multi-object spectroscopy of 42 galaxies from the POWIR survey (Conselice et al. 2008a) with stellar masses M$_{*} > 10^{10}$ and at redshifts $0.4< z <1.3$.  We measure H$\alpha$ luminosities for our sample to calculate their SFRs and to investigate whether the observed trends in global and individual star formation properties are seen when using different SFR tracers.  In this paper we furthermore investigate the number of galaxies with H$\alpha$ emission as a function of mass, compare H$\alpha$-derived SFRs with other measures of their SFRs and relations between the SFRs, and mass and colour.
    
    This paper is organised as follows.  The data and new observations are summarised in Section \ref{sec:Data}, including different methods of correcting for dust attenuation in Section \ref{sec:Corrections}. The galaxies where H$\alpha$ emission is detected are investigated in Section \ref{sec:Results} and the SFR relations are investigated in Section \ref{sec:StarFormation}.  The results, and the astrophysical interpretation of the SFRs are discussed in Section \ref{sec:Discussion}, with a summary of the work in Section \ref{sec:Conclusion}.  Throughout the paper, we adopt the cosmology $\Omega_{m}=0.3$, $\Omega_{\Lambda}=0.7$ and $H_{0}=70$\,km\,s$^{-1}$\,Mpc$^{-1}$.

  \section{Data \& sample}
    \label{sec:Data}
      \subsection{Sample}
      The galaxies investigated in this paper are selected from galaxies within the POWIR survey (Conselice et al. 2008a) as investigated in Conselice et al. (2007, hereafter C07).  The POWIR survey is a large area ($\sim$1.5\,deg$^2$) deep NIR survey in the \textit{K} and \textit{J} band covering the GOODS field North (Giavalisco et al. 2004), the Extended Groth Strip (EGS; Davis et al. 2007) and three fields observed by the DEEP2 team with the DEIMOS spectrograph (Davies et al. 2003).  C07 used all the areas covered except the GOODS-North field.
      
      Data in C07 was acquired from the Palomar 5-m telescope ($K_s$ and $J$ band imaging) and used imaging from the \textit{Canada-France-Hawaii Telescope} (CFHT; B, R, I), the Advanced Camera for Surveys (ACS) on the \HST (Optical), the Multiband Imaging Photometer (MIPS) on the \textit{Spitzer Space Telescope} (24\,$\mu$m), the \textit{Very Large Array} (1.4 GHz), the \textit{Chandra X-ray Observatory} and spectra from the Deep Imaging Multi-Object Spectrograph (DEIMOS) at the W. M. Keck Observatory ($\lambda = 6500-9100$\AA).  A $K$-selected catalogue was created and the different photometry bands matched.  Readers are directed to C07 for more information on these observations and the parent sample selection.
      
      The C07 catalogue was used to measure the stellar masses of galaxies to $K\sim$ 20 -- 21 from SED fitting as in Bundy et al. (2006; for details, see C07).  This measurement uses a Chabrier initial mass function (hereafter IMF; Chabrier 2003).  A complete subset was then created with M$_{*}>10^{10}$ \msol and a redshift range of $z \sim$ 0.4 -- 1.4 taken from DEEP2 spectroscopic redshifts (Davis et al. 2003).  This is the dataset from which our sample was constructed.
      
      The targets for \LIRIS were taken from two regions, one purely mass selected from the POWIR Field 3 to have M$_{*}>10^{10}$\,\msol, the second selected from the EGS field.  This second set of targets were selected to be massive (M$_{*}>10^{10}$\,\msol), have a significant MIPS flux and no Chandra 2-8 keV X-ray emission at $<10^{42}$\,erg\,s$^{-1}$\,cm$^{-2}$.  Both sets were selected within the range $0.4<\mbox{z}_{\rm{spec}}<1.3$, giving a total of $\sim$7500 possible targets.  Masks for multi-object spectroscopy were created using pointings that maximised the number of galaxies targeted.
      
      \subsection{NIR Spectra}
      Our Near infrared (NIR) spectra were taken during 9 nights over two observing runs on the 4.2-metre \WHT\ using the Long-slit Intermediate Resolution Infrared Spectrograph (\textit{LIRIS}) instrument in Multi-Object Spectroscopy mode (Acosta-Pulido et al. 2003).  The spectra were taken using a 3-point nodding routine and the total integration times range from 2.1 hrs to 6.3 hrs for each of the 7 masks.  Each mask targeted between 8 and 14 galaxies, targetting a total of 71 galaxies.
      
\begin{figure}
        \centering
        \includegraphics[angle=0, width=8.5cm]{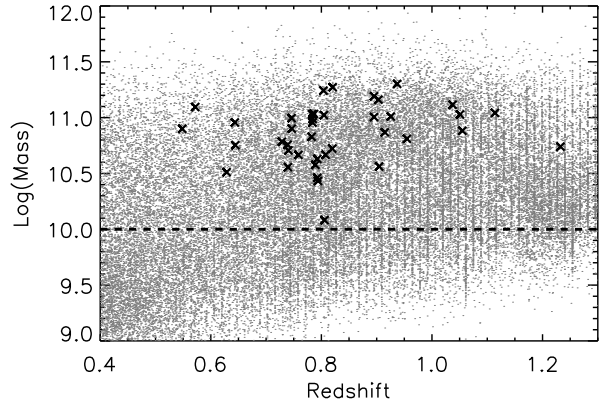}
        \caption{The redshifts and stellar masses of the POWIR survey (grey points) and our observed galaxies taken from this sample (black triangles).  The dashed line is our mass cut.}
        \label{fig:MassZ}
      \end{figure}

      Data reduction was done using the dedicated suite of routines {\sc lirisdr} for {\sc iraf}, written by Jos\'{e} Acosta-Pulido and the spectra were flux calibrated using $J$-band magnitudes from the POWIR survey data.  In all cases the spectra covered the wavelength range of the $J$-band, and the spectra were scaled to give the same integrated flux measurement within that wavelength range as the $J$-band magnitude measurements.  This also corrected for flux lost due to the finite width of the slit (aperture correction).  Out of the 71 galaxies targeted, 42 had detected continuum.  There are no galaxies where there is emission at around 6563 \AA\ but have no detected continuum.  There were 21 galaxies detected in each of the two areas of sky targeted.  One galaxy, 32014349 is affected by an atmospheric absorption feature within 200\AA\ of the wavelength of any H$\alpha$ emission line and so is discarded leaving 41 galaxies in our sample.  The stellar mass and redshifts of these 41 galaxies are shown in Figure \ref{fig:MassZ} (see also Table \ref{tab:data}).
      
      The 1D Spectra were examined by eye and if an emission line was seen at the correct wavelength to be H$\alpha$, the flux was extracted from the continuum subtracted spectra within this region.  Even though in most cases where there was substantial $H\alpha$ emission the H$\alpha\lambda6563$ and [NII]$\lambda6583$ appeared separated, the extraction region was chosen to include both emission lines.  This was to allow the procedure for measuring an upper limit where there was no emission detected to follow as closely as possible the procedure used when there was emission.  The continuum level and noise was taken from two 200\,\AA\ sections of continuum one average full width half maximum (FWHM, 5 pixels = 31\,\AA) either side of the emission and the root mean squared value of this continuum is used as the error on each pixel.  This average FWHM is used as the extraction width for continuum with no visual signs of emission.
      
      If the H$\alpha$ emission is less than the 1.5 times the error and no line has been seen by eye, then these galaxies are classified as non-detections.  This method of measuring the emission line is similar to that used in Glazebrook et al (1999).  A correction of 30 per cent for [NII] is applied (see for example, Gallego et al. 1997; Tresse et al. 1999; Garn et al. 2010).  In cases where there was no visible emission, the continuum noise was used to calculate a 1.5 $\sigma$ upper limit to any H$\alpha$ emission.  The DEEP2 spectroscopic redshifts were used to determine the predicted observed H$\alpha$ wavelengths and convert fluxes to luminosities.

      \subsection{Star formation rates}
      Star formation rates were derived from the H$\alpha$ luminosities using the Kennicutt (1998) relation converted to use a Chabrier IMF (Chabrier 2003):
      
      \begin{equation}
        \mbox{SFR}\left(\mbox{\msol\,yr}^{-1}\right) = 4.6\times10^{-42}L(H\alpha)\,\mbox{(erg\,s$^{-1}$)}.
        \label{eqn:K98Ha}
      \end{equation}
      
      \noindent In order to compare our measurements with measurements taken in other wavelengths, we use Infrared (IR), UV and radio observations already available of the same sources.  Ultraviolet (UV) data was taken using the \textit{Galaxy Evolution Explorer} (GALEX) and available imaging in the $B$, $R$ and $I$ filters to match the galaxy's rest-frame emission as close as possible to the GALEX filters (1528\,\AA\ and 2271\,\AA; Schiminovich 2007).  SFRs and dust extinction-corrected SFRs were calculated using a composite method using IR calibrated UV-optical colours and broadband SED fitting (for a description of the method see Schiminovich et al. 2007 and Seibert et al. 2005).  The UV and IR SFRs are discussed in more detail in Section \ref{sec:Corrections}.
    
      Radio data was acquired from a 1.4 GHz catalogue of the EGS, (Ivison et al. 2007) taken with the \textit{Very Large Array} (VLA).  They detect 1,123 sources with S$_{1.4\mbox{\,GHz}} \geq 50$\,$\mu$Jy to a 5-$\sigma$ detection limit which corresponds to rest-frame 1.4 GHz calculated SFR of 275\,\msol\,yr$^{-1}$ at z = 1.0.
    
      Our sample were matched with the radio catalogue, but only reveal 1 match (from 41 galaxies).  The radio catalogue was then matched with the entire set of possible galaxies that could have been observed and are in the same region of sky as those that were observed.  Out of these 62 galaxies there are 4 matches ($6\pm3$ per cent).
      
      Concluding that our galaxies are not detected at 1.4 GHz, upper limits can be placed on the total star formation in these galaxies following the methods of Condon (1992), Haarsma et al. (2000) and Condon, Cotton \& Broderic (2002).  Using the detection limit of the nearest galaxy in the sample (z $\sim$ 0.55) gives an upper limit of 66\,\msol\,yr$^{-1}$, the furthest (z $\sim$ 1.23) gives $\sim$450\,\msol\,yr$^{-1}$.  These are considerably higher than the majority of our upper limits derived from the H$\alpha$ measurements and thus are not further used.
      
      \begin{table*}
        \begin{center}
          \begin{tabular}{c c c c c c c c c c}
            \hline
            ID & z & $(U-B)$ & $\log$M$_{*}$ & L$_{\rm{H\alpha}}$                   & $E(B-V)$ & SFR$_{\rm{H\alpha,uncorr}}$  & SFR$_{\rm{H\alpha,corr}}$ & SFR$_{24\,\mu m}$   & SFR$_{\rm{UV}}$ \\ 
               &   &         & (\msol) & ($\times 10^{41}$\,erg\,s$^{-1}$) &          & (\msol\,yr$^{-1}$) & (\msol\,yr$^{-1}$) & (\msol\,yr$^{-1}$) & (\msol\,yr$^{-1}$) \\\hline

12019914 & 0.57 & 1.21 & 11.10 & $<$2.5 & 0.40 & $<$2.0 & $<$5.1 & 10.9 & 11.1 \\
12019916 & 0.73 & 0.84 & 10.78 & $<$6.8 & 0.39 & $<$5.4 & $<$13.3 & 22.2 & 20.3 \\
12020027 & 0.63 & 0.77 & 10.51 & 19.9 $\pm$ 4.0 & 0.45 & 15.7 $\pm$ 3.1 & 45.1 $\pm$ 54.1 & 8.7 & 36.4 \\
12020031 & 0.82 & 0.97 & 11.27 & $<$19.4 & 0.30 & $<$15.4 & $<$31.0 & 15.6 & 66.4 \\
12020035 & 0.55 & 0.92 & 10.90 & 1.7 $\pm$ 1.1 & 0.35 & 1.4 $\pm$ 0.9 & 3.1 $\pm$ 5.0 & 7.2 & 16.2 \\
12024309 & 0.64 & 0.86 & 10.96 & 12.2 $\pm$ 3.1 & 0.25 & 9.7 $\pm$ 2.4 & 17.5 $\pm$ 21.8 & 19.6 & 23.6 \\
12024315 & 1.05 &  --  & 11.03 & $<$9.3 & 0.05 & $<$7.3 & $<$8.2 & 9.9 & 8.5 \\
12024359 & 0.74 &  --  & 10.76 & $<$10.3 & 0.24 & $<$8.1 & $<$14.1 & -- & 4.2 \\
12024380 & 0.76 & 0.97 & 10.67 & $<$4.9 & 0.62 & $<$3.9 & $<$16.5 & 11.6 & -- \\
12024427 & 0.93 & 1.08 & 11.01 & $<$1.9 & 0.31 & $<$1.5 & $<$3.1 & 13.6 & 24.0 \\
12024436 & 0.94 & 0.92 & 11.30 & 29.5 $\pm$ 8.6 & 0.42 & 23.3 $\pm$ 6.8 & 61.5 $\pm$ 79.5 & 84.5 & 128.1 \\
12024440 & 0.64 & 0.90 & 10.75 & 5.9 $\pm$ 3.0 & 0.26 & 4.7 $\pm$ 2.4 & 8.6 $\pm$ 13.0 & 4.2 & 11.2 \\
12024445 & 0.91 & 0.85 & 10.87 & 39.7 $\pm$ 5.4 & 0.38 & 31.4 $\pm$ 4.3 & 75.5 $\pm$ 85.7 & 26.1 & 99.4 \\
12024528 & 0.90 & 0.72 & 10.56 & 12.4 $\pm$ 3.3 & 0.34 & 9.8 $\pm$ 2.6 & 21.7 $\pm$ 27.4 & 14.8 & 29.5 \\
13026831 & 1.11 & 0.99 & 11.04 & $<$16.0 & 0.66 & $<$12.6 & $<$59.5 & 24.7 & -- \\
13027239 & 1.23 & 0.87 & 10.74 & $<$4.0 & 0.61 & $<$3.1 & $<$13.2 & -- & -- \\
13027461 & 0.81 & 0.97 & 10.67 & $<$2.3 & 0.61 & $<$1.8 & $<$7.4 & 23.3 & -- \\
13035323 & 0.75 & 0.99 & 11.00 & 15.5 $\pm$ 5.0 & 0.66 & 12.3 $\pm$ 3.9 & 57.7 $\pm$ 76.2 & 24.4 & -- \\
13035329 & 0.75 & 1.12 & 10.90 & $<$4.3 & 0.62 & $<$3.4 & $<$14.1 & 20.1 & -- \\
13035770 & 0.95 & 0.83 & 10.81 & $<$8.7 & 0.63 & $<$6.8 & $<$30.1 & -- & -- \\
13035775 & 1.06 & 0.95 & 10.88 & $<$2.1 & 0.61 & $<$1.7 & $<$6.9 & 7.8 & -- \\
32003256 & 0.90 & 1.05 & 11.16 & $<$5.5 & 0.62 & $<$4.4 & $<$18.6 & -- & -- \\
32008396 & 0.80 & 1.16 & 11.24 & $<$3.1 & 0.61 & $<$2.4 & $<$10.1 & -- & -- \\
32008735 & 0.90 & 1.24 & 11.19 & $<$2.3 & 0.61 & $<$1.8 & $<$7.5 & -- & -- \\
32008739 & 0.79 & 1.07 & 11.03 & $<$2.6 & 0.61 & $<$2.0 & $<$8.5 & -- & -- \\
32008871 & 1.04 & 1.09 & 11.12 & 28.8 $\pm$ 3.7 & 0.72 & 22.8 $\pm$ 2.9 & 121.7 $\pm$ 137.2 & -- & -- \\
32008909 & 0.79 & 1.01 & 10.58 & $<$0.9 & 0.60 & $<$0.7 & $<$3.0 & -- & -- \\
32008915 & 0.74 & 0.75 & 10.56 & 2.1 $\pm$ 0.9 & 0.61 & 1.6 $\pm$ 0.7 & 6.7 $\pm$ 9.7 & -- & -- \\
32008924 & 0.78 & 0.83 & 10.96 & 6.5 $\pm$ 1.5 & 0.63 & 5.1 $\pm$ 1.2 & 22.1 $\pm$ 27.2 & -- & -- \\
32009330 & 0.90 & 1.05 & 11.01 & $<$4.2 & 0.62 & $<$3.3 & $<$13.9 & -- & -- \\
32014288 & 0.78 & 1.03 & 10.83 & $<$3.9 & 0.61 & $<$3.0 & $<$12.8 & -- & -- \\
32014399 & 0.79 & 0.66 & 10.63 & 50.9 $\pm$ 3.6 & 0.81 & 40.2 $\pm$ 2.9 & 266.7 $\pm$ 285.7 & -- & -- \\
32014580 & 0.79 & 0.61 & 10.46 & 15.6 $\pm$ 2.4 & 0.66 & 12.3 $\pm$ 1.9 & 57.9 $\pm$ 66.9 & -- & -- \\
32018834 & 0.81 & 0.59 & 10.08 & 2.9 $\pm$ 1.3 & 0.61 & 2.3 $\pm$ 1.0 & 9.5 $\pm$ 13.7 & -- & -- \\
32018902 & 0.82 & 1.29 & 10.72 & $<$7.3 & 0.63 & $<$5.8 & $<$25.0 & -- & -- \\
32018960 & 0.78 & 1.14 & 11.03 & $<$18.5 & 0.68 & $<$14.6 & $<$70.7 & -- & -- \\
32019661 & 0.78 & 1.02 & 11.03 & $<$3.0 & 0.61 & $<$2.3 & $<$9.7 & -- & -- \\
32019921 & 0.78 & 1.17 & 10.99 & $<$5.8 & 0.62 & $<$4.6 & $<$19.6 & -- & -- \\
32019951 & 0.80 & 1.05 & 11.02 & $<$9.7 & 0.64 & $<$7.6 & $<$33.8 & -- & -- \\
32024002 & 0.74 & 1.31 & 10.71 & $<$11.5 & 0.65 & $<$9.1 & $<$41.2 & -- & -- \\
32024419 & 0.79 & 0.65 & 10.43 & $<$11.3 & 0.64 & $<$8.9 & $<$40.2 & -- & -- \\

\hline

          \end{tabular}
        \end{center}
        \caption{Galaxies observed in our sample.  $(U-B)$ is colour, $\log$M$_{*}$ is the stellar mass, L$_{\rm{H\alpha}}$ is the H$\alpha$ luminosity, $E(B-V)$ is the $(B-V)$ colour excess, SFR$_{\rm{H\alpha,uncorr}}$, SFR$_{\rm{H\alpha,corr}}$, SFR$_{24\,\mu m}$ and SFR$_{\rm{UV}}$ are the star formation rates measured via H$\alpha$ emission uncorrected and corrected for dust, 24-$\mu$m emission and UV emission.  The H$\alpha$ derived SFR is not corrected for dust, the UV derived SFR has been corrected.}
        \label{tab:data}
      \end{table*}

      \subsection{Imaging}
      \label{sec:Imaging}
\begin{figure*}
        \centering
	\includegraphics[angle=0]{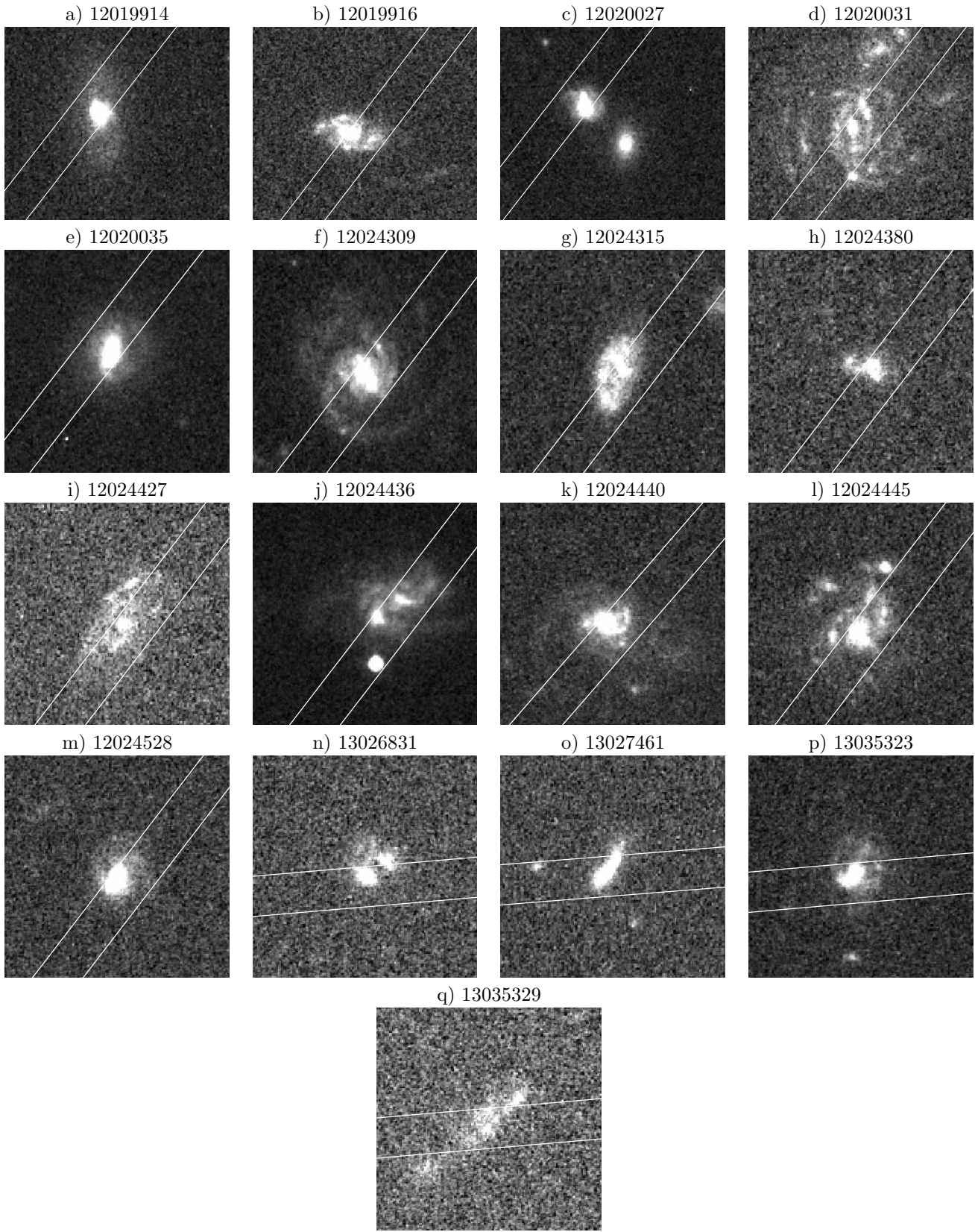}

        \caption{\HST \textit{ACS} F814W images of the galaxies where HST data is available in the EGS field (Conselice et al. 2007) overlaid with the slit position and width (white lines).  The field of view of each image 9'' by 9''.}
        \label{fig:Images}
      \end{figure*}

      Figure \ref{fig:Images} shows the subset of our sample imaged by the \HST \textit{ACS} camera through the F606W filter with the position of the \LIRIS slits overlaid.  
      In general, the slits are well aligned with the target galaxies.  
      For all galaxies, the slit captures the centre and some of the outer-regions and so we can be confident that we are capturing a reasonable representation of the emission from star formation.
      
      The majority of the galaxies in our sample are clearly disk galaxies.  There are a couple of galaxies with disturbed morphologies and signs of possibly being merging systems.  Note that none of the galaxies in our sample were selected by morphology.  Below we list some features of our four systems whose observations may be contaminated by nearby galaxies or stars.
      
      Galaxy 12020027 (Figure \ref{fig:Images}c) is actually two galaxies separated by 1.4''.  These two galaxies are close enough together that the 24-$\mu$m infrared data from the \textit{Spitzer Space Telescope} will not differentiate between the two galaxies due to its large point spread function (PSF).  If the second galaxy has a higher dust content than the target galaxy, the H$\alpha$ emission will be dominated by the galaxy with the least dust, whereas the 24-$\mu$m emission is dominated by emission from the dusty galaxy.  This will lead to an applied dust correction that is too large (or a too large amount of H$\alpha$ emission being corrected).  This is a possible explanation for the high dust-corrected SFR from this galaxy (see Section \ref{sec:Corrections}).
      
      Galaxy 12020031 (Figure \ref{fig:Images}d) has a possible second galaxy 2.0'' away from the centre of the main galaxy.  Again this may affect the dust-corrected SFR measured.
      
      Galaxy 12024436 (Figure \ref{fig:Images}j) appears to have a small object, probably a foreground star also in the slit, 1.5'' away from the galaxy.  This is not an obvious extra component in the spectra, but may affect other measurements of the galaxy.  This galaxy has the appearance of a merging system with two central peaks separated by 0.6''
      
      Galaxy 13026831 (Figure \ref{fig:Images}n) also appears to have two parts separated again by 0.6''

    \subsection{Dust corrections of star formation rates}
    \label{sec:Corrections}
    Each method of measuring the star formation rate (SFR) of galaxies uses emission at certain wavelengths.  Optical tracers of SFR are affected by dust extinction that is a function of wavelength, so we must apply these corrections.  In the NIR, optical, and UV wavelengths, the effect is manifested as an attenuation to the emission that is highly dependent on wavelength (see for example, Calzetti 2001).
    
    Many studies -- usually at redshifts where observations of emission lines becomes difficult, or where spectroscopic data is not yet available -- often use a blanket value for the amount of attenuation a galaxy suffers at a specific wavelength, $A(\lambda)$ (often 1 magnitude at V-band: $A(H\alpha)\sim0.8$) and apply it to all the galaxies in their sample (e.g. Kennicutt 1983; Charlot et al. 2002; Tresse et al. 2002; Garn et al. 2010 and references within).  There is evidence however, that there are a range of attenuations in galaxies at $z \sim1$ (e.g. Garn et al. 2010).  In addition, there is also a large uncertainty of up to $\sim$30 per cent on the conversion of H$\alpha$ luminosity to SFR due to different values derived from different stellar population models (see Kennicutt 1998a).  Therefore this method can crudely work to statistically correct a large sample.  Our sample of galaxies is neither large enough, nor statistically general enough to apply such a method.
    
    There are two sources of emission used at these wavelengths for measuring the SFR -- nebula emission lines (e.g. H$\alpha$, H$\beta$, Pa$\alpha$, OII) and stellar emission (usually in the UV).  Emission lines are caused by recombination or collisional excitation in HII regions which are formed by massive, short-lived O and B type stars; similarly the UV emission gives the equivalent of a number count of the these stars.  Both of these have been calibrated with certain assumptions to give simple conversions to the SFR of the galaxy (see Kennicutt 1998a).
    
    The attenuation of a galaxy at a specific wavelength, $A(\lambda)$, is related to the colour excess of the nebular emission in the galaxy, $E(B-V)_{\rm{gas}}$ as given by:
    
    \begin{equation}
      A(\lambda)=E(B-V)_{\rm{gas}}\,k(\lambda)
    \end{equation}
    
    \noindent where $k(\lambda)$ is the value of the dust extinction curve being used at the wavelength of emission.  There is significant large-scale differences between different dust extinction curves that have been measured to date.  However a study by Calzetti (2001) showed that between the (rest-frame) wavelengths of $\sim$3700\,\AA\ and $\sim$6600\,\AA\ the differences are minor.  The Galactic extinction curve of Cardelli, Clayton \& Mathis (1989) is used in this study which gives $k(H\alpha)=2.54$, $k(H\beta)=3.61$ and $k(H\gamma)=4.17$.
    
    A standard method to correct for dust on a galaxy by galaxy basis is to measure the ratio of the H$\alpha$ to H$\beta$ emission and compare this to the theoretical value (the Balmer decrement).  As these lines will be attenuated by different amounts, a dust extinction curve can be used to calculate the amount of attenuation.  This method can be used with any two emission lines that have a known intrinsic ratio.  Five galaxies in our sample have both H$\alpha$ measurements (not upper limits) from our measurements, and H$\beta$ luminosities from the DEEP2 spectroscopy allowing for a calculation of the dust correction through the Balmer decrement.  Eight galaxies have H$\gamma$ from DEEP2 allowing a comparison with the H$\beta$ lines.  Three galaxies have all three values.  Intrinsic ratios of $(F_{i}^{H\alpha}/F_{i}^{H\beta})=2.87$ and $(F_{i}^{H\gamma}/F_{i}^{H\beta})=0.466$ (Case B recombination at T = 10,000K, Osterbrock 1989) are used to calculate $E(B-V)_{gas}$ values.
    
    
    Even within the large errors, we find significant differences in the range of measured values (Table \ref{tab:hahbhghb}), so that unfortunately this extinction correction method does not give satisfying results.  The large errors are likely the result of insufficient signal-to-noise ratio of both our present H$\alpha$ survey and the DEEP2 survey's H$\beta$ and H$\gamma$ measurements.
    
    \begin{table}
      \begin{center}
        \begin{tabular}{c c c}
          \hline
          ID & \multicolumn{2}{c}{$E(B-V)_{\rm{gas}}$ from} \\
           & H$\alpha/$H$\beta$ & H$\gamma/$H$\beta$ \\ \hline
          12019914 &  --                & 3.8 $\pm$ 1.3    \\
          12024309 & 1.98 $\pm$ 0.43    & -0.54 $\pm$ 0.64 \\
          12024427 &  --                & 2.60 $\pm$ 0.72   \\
          12024440 & 0.83 $\pm$ 0.69    & 0.83 $\pm$ 0.99  \\
          12024445 & 3.8 $\pm$ 1.6      &  --                \\
          12024528 & 0.91 $\pm$ 0.49    &  --                \\
          13027461 &  --                & 0.30 $\pm$ 20   \\
          13035323 & -0.08 $\pm$ 0.50 & 0.57 $\pm$ 0.22  \\
          13035770 &  --                & 1.59 $\pm$ 0.60    \\
          13035775 &  --                & 2.1 $\pm$ 1.1      \\
          \hline
        \end{tabular}
      \end{center}
      \caption{$E(B-V)_{\rm{gas}}$ values derived from the Balmer decrements for the galaxies where line fluxes are available.  Values would be expected to be $E(B-V)_{\rm{gas}} \lesssim 1.5$; negative values are unphysical}
      \label{tab:hahbhghb}
    \end{table}
    
    We can also use emission in the far infrared to calculate dust corrections, as UV light absorbed by dust is re-radiated in the far-IR and this can be used to correct the attenuated SFRs measured in other wavelengths.  Data taken from the \Spitzer MIPS camera was $K$-corrected and converted to SFRs by Noeske et al. (2007).  As this is a measure of obscured SFR, this data can then be combined with UV data to correct for dust.
    
    We use two methods to account for dust.  A direct but simplistic method to obtain the total SFR is to use a summation of the un-obscured SFR from the UV and the obscured SFR from the 24$\mu$m.
    
    \begin{equation}
      \mbox{SFR}_{\rm{Tot}} = \mbox{SFR}_{\rm{UV}} + \mbox{SFR}_{24\,\mu m},
      \label{eqn:SFRtot}
    \end{equation}
    
    \noindent where, when calculating the SFR$_{\rm{UV}}$ from the far-UV (FUV) flux (flux detected at the wavelength of the \GALEX FUV band which has an effective wavelength of 1528\,\AA)
    
    \begin{equation}
      \mbox{SFR}_{\rm{UV}}\left(\mbox{\msol\,yr}^{-1}\right) = 10^{-28.165}L_{\rm{FUV}}\mbox{(erg\,s$^{-1}$\,Hz$^{-1}$)},
          \label{eqn:SFRUV}
    \end{equation}
    
    \noindent (Salim et al. 2007) and the SFR calculated from the IR emission
    
    \begin{multline}
      \mbox{SFR}_{24\,\mu m}\left(\mbox{\msol\,yr}^{-1}\right)=\\
          1.27\times10^{-38}\left[L_{24\,\mu m}\mbox{(erg\,s$^{-1}$)}\right]^{0.8850},
    \end{multline}
    
    \noindent (Calzetti et al. 2007).  This requires accurate calibrations of both the derived SFRs, and ignores possible effects from dust geometry and complex radiative transfer processes.  Kennicutt (2009) derive a set of coefficients that are multiplied by an IR SFR tracer luminosity to convert it into the luminosity from an attenuated emission-line SFR tracer that has been absorbed by dust.  In both of these methods the IR SFR (luminosity) is much larger than the H$\alpha$ SFR (luminosity) and therefore the corrected SFR becomes approximately the IR SFR
    
    
    Schiminovich et al. (2007) developed a method for correcting the UV-derived SFRs for dust attenuation using a hybrid approach combining an IR-calibrated measure of the far-UV (FUV) attenuation ($A_{IRX}$) based on UV-optical colours and the 4000\AA\ break $D_{n}(4000)$ from Johnson et al. (2006, 2007) (accurate for galaxies with $D_{n}(4000)<1.7$) and an attenuation measure $A_{z}$ that is based on detailed model fits to the SDSS absorption-line spectrum and broadband SED from Kauffmann et al. (2003a) (likely to be more accurate for galaxies with higher $D_{n}(4000)$).  The combined fit is used with the Calzetti et al. (2000) extinction curve to correct the UV luminosity, and then the relation from Salim et al. (2007, Equation \ref{eqn:SFRUV}) to derive the SFR.  For more details and discussion, see Schiminovich et al. (2007).  We find that 13 out of our sample of 41 galaxies have SFRs that can be derived from \GALEX UV data.
    
    Using either of these methods, we can calculate the total amount of star formation in each galaxy.  This can then be compared with the SFR measured using H$\alpha$ emission to determine the amount of extinction in the H$\alpha$ flux.  As the UV emission comes directly from stars within a galaxy, the ratio of detected UV emission to dust corrected UV emission can be used to measure the $E(B-V)$ value for the stellar component of the galaxy.  This $E(B-V)_{\rm{stellar}}$ value must be converted to the $E(B-V)_{\rm{gas}}$ value for use with nebular emission lines.  The relationship has been empirically measured as
    
    \begin{equation}
      E(B-V)_{\rm{stellar}}=(0.44\pm 0.03)\,E(B-V)_{\rm{gas}}.
      \label{eqn:E_BVstellar}
    \end{equation}
    
    \noindent (Calzetti, Kinney \& Storchi-Bergmann 1994; Calzetti et al. 2000).
    
     There appears to be a correlation between a galaxy's dust content and its SFR (e.g. Hopkins et al. 2001; Sullivan et al. 2001; Bauer et al. 2011).  This may arise from the fact that the SFR in a galaxy is proportional to the gas surface density (Kennicutt 1998b) and so galaxies with higher SFRs must have more gas and dust and hence more extinction.  From this idea, many studies have tried to create either a theoretical or empirical law that relates the amount of star formation in a galaxy to the dust extinction.  Three such relations are described by Afonso et al. (2003), Choi et al. (2006) and Garn et al. (2010).
     
    
    \begin{figure*}
      \centering
      \includegraphics[angle=0]{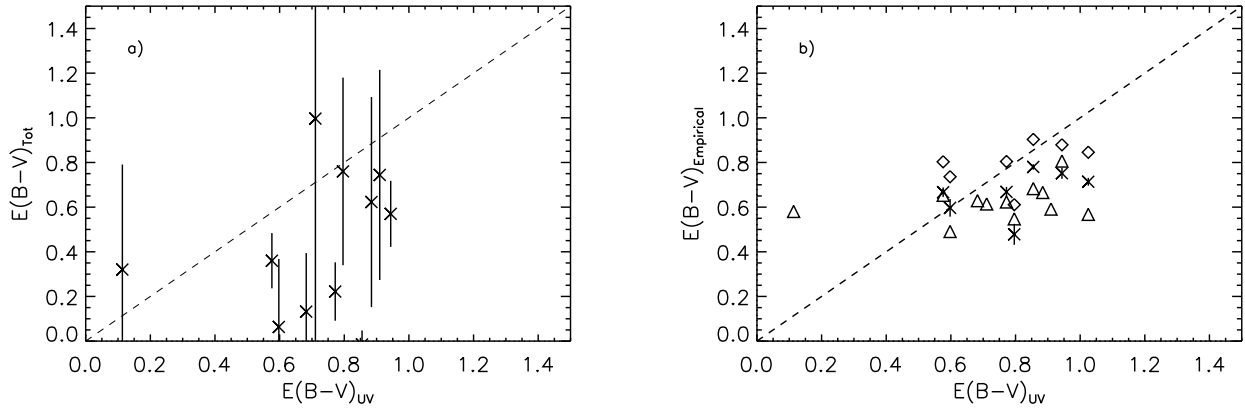}
      \caption{a) Comparison of the UV-derived $E(B-V)_{\rm{gas}}$ values ($E(B-V)_{\rm{UV}}$) with the $E(B-V)_{\rm{gas}}$ values derived from the total UV+IR SFR ($E(B-V)_{\rm{Tot}}$).  The dashed line is the 1:1 relationship.  b) Comparison of the UV-derived $E(B-V)_{\rm{gas}}$ values with the empirically derived values ($E(B-V)_{\rm{Empirical}}$) from relations in Garn et al. (2010, crosses), Choi et al. (2006, triangles) and Afonso et al. (2003, diamonds).  The relations in Choi et al. and Afonso et al. have been converted from a Salpeter IMF to a Chabrier IMF by lowing the SFR of each point by a factor of 1.7.}
      \label{fig:CompE_BV}
    \end{figure*}
    
    \begin{figure}
      \centering
      \includegraphics[angle=0, width=8.5cm]{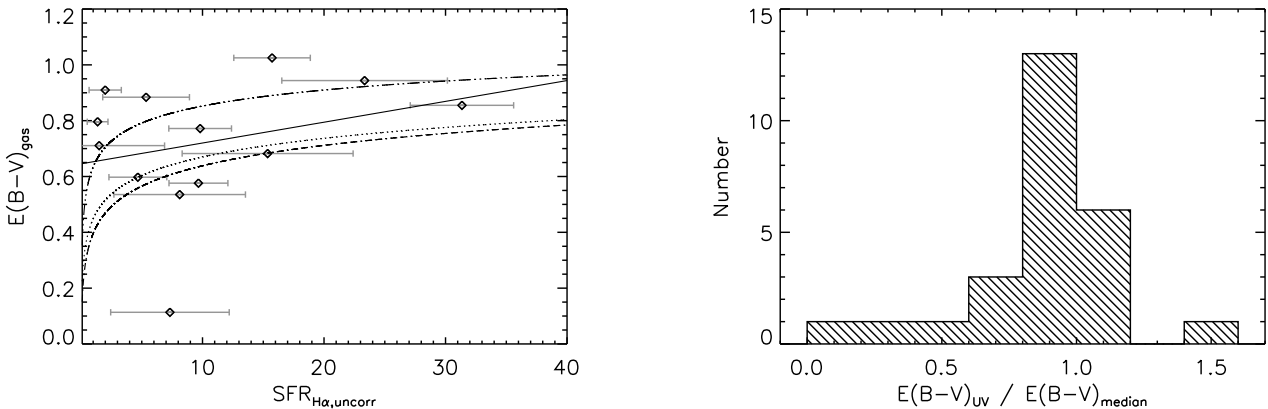}
      \caption{Comparison of the $E(B-V)_{\rm{gas}}$ values derived from UV measurements as a function of uncorrected SFR with the empirically derived relations shown.  The solid line shows the best fit relation from the UV-derived values.  The three curved lines are the empirical models, Afonso et al. 2003 (dot-dot-dash), Choi et al. 2006 (dot-dash), Garn et al. 2010 (dotted).}
      \label{fig:E_BVSFR}
    \end{figure}
    
    Figure \ref{fig:CompE_BV} shows the comparison between $E(B-V)_{\rm{gas}}$ correction values derived from the UV data (Equation \ref{eqn:SFRUV}) with the corresponding $E(B-V)_{\rm{gas}}$ values derived from the total SFR (Equation \ref{eqn:SFRtot}) and with the $E(B-V)_{\rm{gas}}$ values from the empirical relations of Alfonso et al., Choi et al., and Garn et al.  The $E(B-V)_{\rm{gas}}$ values derived from the UV are consistent with the values derived from the empirical methods.  Figure \ref{fig:E_BVSFR} shows the UV-derived $E(B-V)_{\rm{gas}}$ as a function of observed observed SFR$_{H\alpha}$ along with the empirical relations.  There is a slight increase in $E(B-V)_{\rm{gas}}$ with higher SFR, and this can be fit with a straight line by a least-square method function as:
    
    \begin{multline}
      E(B-V)_{\rm{UV}} = \\
        \left(8.96\pm13.7\right)\times10^{-3} ~ \mbox{SFR}_{\rm{H\alpha,uncorr}} + \left(0.663\pm0.113\right),
      \label{eqn:UVcorr}
    \end{multline}
    
    \noindent Although with significant scatter.  Each method for deriving the $E(B-V)_{\rm{gas}}$ value for each galaxy has its own biases, short comings and advantages, and as there are such differences between the methods, they cannot be interchanged.
    
    The median $E(B-V)_{\rm{gas}}$ value for each galaxy is calculated from the $E(B-V)$ values derived from the different available methods.  However because not all the methods can be used for all the galaxies, the median value of $E(B-V)$ for each galaxy will not be comparable across our sample.  Some galaxies do not have enough data for any corrections, and where galaxies only have enough data for one method, it then introduces that method's specific biases onto that galaxy only.  Picking only one method ensures the method based biases for all galaxies are the same and facilitates some attempt at correcting for this.
    
    \begin{figure}
      \centering
      \includegraphics[angle=0, width=8.5cm]{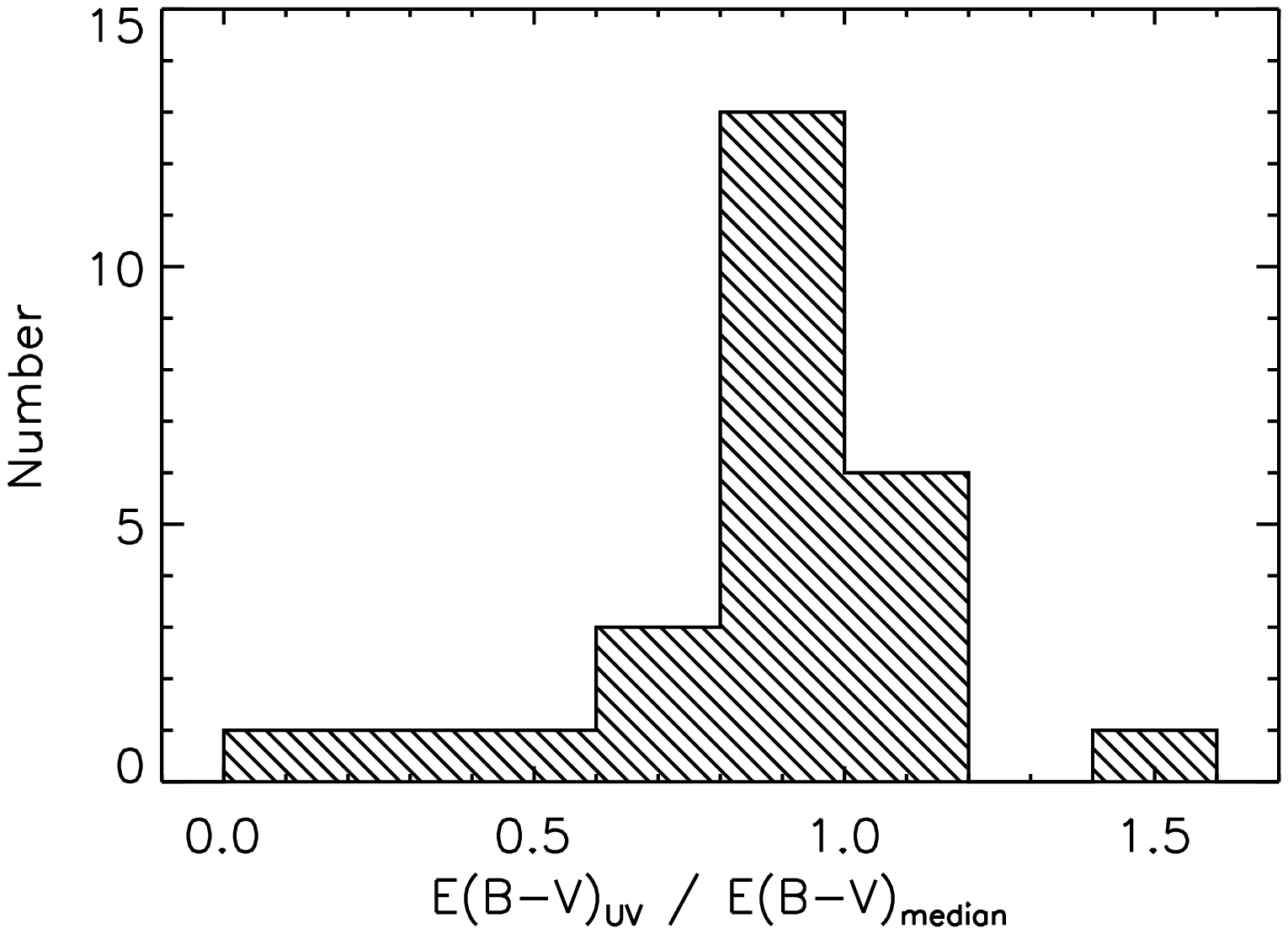}
      \caption{Comparisons of the $E(B-V)_{\rm{gas}}$ values derived from the UV SFRs with the median value of all methods.  Most UV-derived values lie near the median value, or are lower}
      \label{fig:E_BVUVMed}
    \end{figure}
    
    We thus use the $E(B-V)_{\rm{gas}}$ value derived from the \GALEX UV data to correct our H$\alpha$-derived SFRs which allows the most galaxies to be corrected with their own observational data.  The fitted relation in Equation \ref{eqn:UVcorr} is used to correct galaxies for dust extinction with no UV derived SFRs, which equates to an attenuation that ranges from $A_{\rm{H\alpha}}=1.68$ at SFR$_{\rm{H\alpha,uncorr}}=0.1$\,\msol\,yr$^{-1}$ to $A_{\rm{H\alpha}}=2.25$ at SFR$_{\rm{H\alpha,uncorr}}=25$\,\msol\,yr$^{-1}$.  Figure \ref{fig:E_BVUVMed} shows the deviation of the UV-derived $E(B-V)_{\rm{gas}}$ values from the median $E(B-V)_{\rm{gas}}$ value for that galaxy calculated over all our methods.  The UV-derived $E(B-V)_{\rm{gas}}$ values are seen to be in reasonably good agreement with the median values.

  \section{Results}
    \label{sec:Results}

    \subsection{H$\alpha$ detections}
      
      \begin{figure*}
        \centering
        \includegraphics[angle=0]{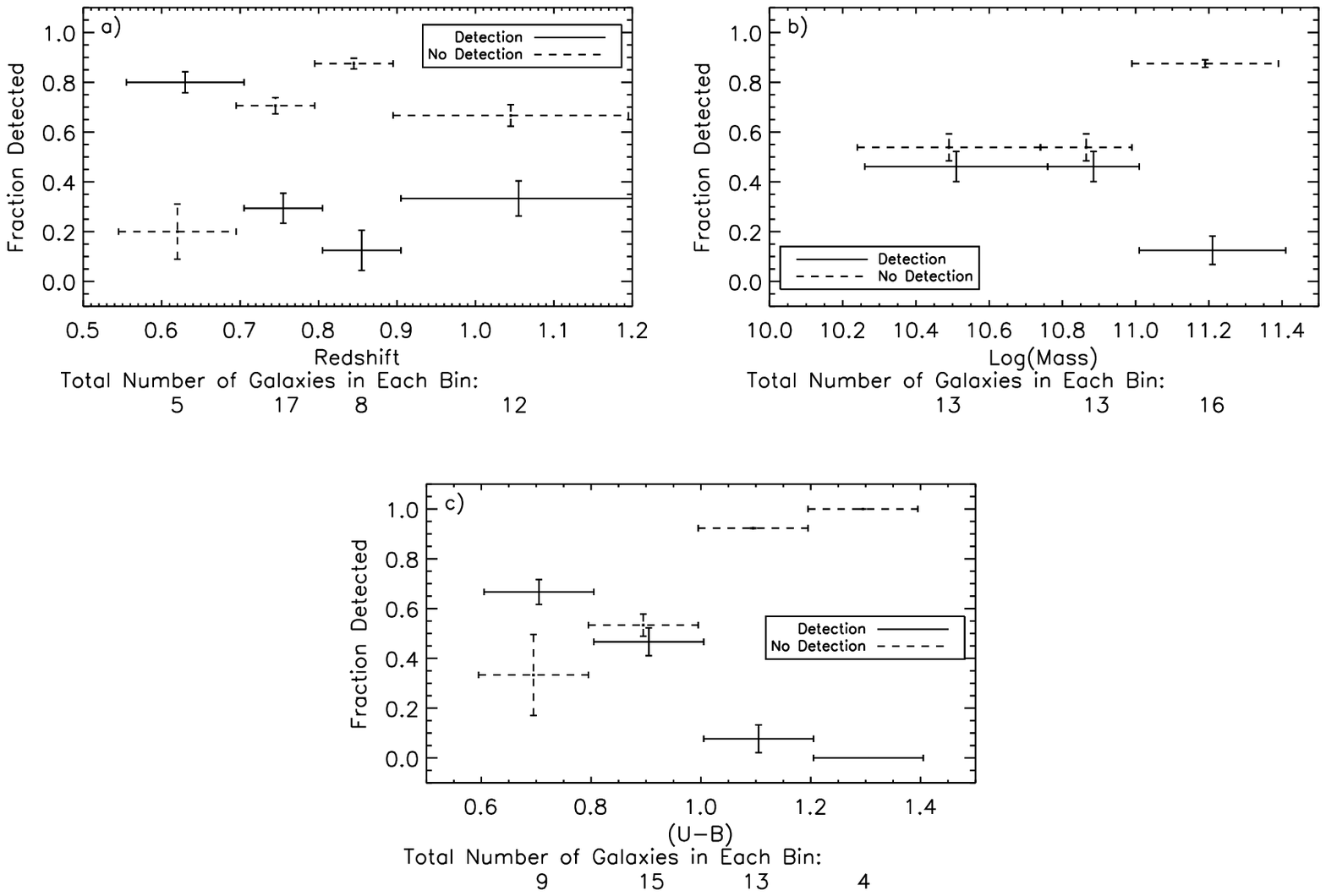}
        
        \caption{a) Fraction of galaxies detected in H$\alpha$ as a function of redshift.  There is no obvious bias against detecting H$\alpha$ emission at high redshift in our study.  b) Fraction of all our galaxies detected in H$\alpha$ as a function of stellar mass.  Galaxies with stellar masses greater than M$_{*}=10^{11}$\,\msol have a significantly lower fraction of H$\alpha$ detections than lower masses.  c) Fraction of galaxies detected in H$\alpha$ as a function of U-B colour.  There is a significant drop in H$\alpha$ detections towards redder galaxies.  The points have been shifted by $\pm 0.005$, $\pm 0.01$ and $\pm 0.005$ respectively on the x-axis of each graph for clarity.}  
        \label{fig:Detections}
      \end{figure*}
      
      \begin{figure}
        \centering
        \includegraphics[angle=0, width=8.5cm]{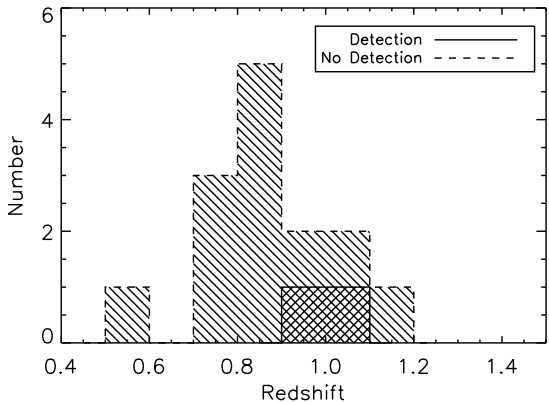}
        \caption{The redshifts of galaxies with M$_{*}>10^{11}$ \msol divided into those with H$\alpha$ detections and those without.  All detections of star formation in these massive galaxies are at $z >0.93$}
        \label{fig:MassiveZ}
      \end{figure}
      
      \label{sec:Detections}
      Out of the 41 galaxies with a continuum detection in the sample, 14 ($34\pm10$ per cent) have significant detected H$\alpha$ emission.  The median detection limit for uncorrected star formation is $3.0\pm0.5$\,\msol\,yr$^{-1}$.  First, to investigate which type of galaxies are being detected, the fraction of detections are plotted as a function of redshift in Figure \ref{fig:Detections}a.  At redshifts higher than $z$ = 0.7, there is no statistical change in the detection rate as a function of redshift, implying there is no strong bias against galaxies at either end of our redshift range in our sample.
      
      The errors plotted for each point are the combination of the error on the number of galaxies detected in H$\alpha$, $\sigma_{N_{d}}$ and the total number of galaxies, $\sigma_{N_{T}}$, assuming Poisson statistics for both, with a cross term that takes into account the fact that these two quantities are not independent.
      
      \begin{equation}
        \frac{\sigma_{f}^{2}}{f^{2}} \simeq \frac{\sigma_{N_{d}}^{2}}{N_{d}^{2}} + \frac{\sigma_{N_{T}}^{2}}{N_{T}^{2}} - \frac{2 \sigma_{N_{d}N_{T}}^{2}}{N_{d}N_{T}},
      \end{equation}
      
      \noindent (Bevington 1969). $\sigma_{N_{d}N_{T}}^{2}$ is the covariance between $N_{d}$ and $N_{T}$ and for a positively-correlated pair of gaussian-distributed sets, can be approximated as
      
      \begin{equation}
        \sigma_{N_{d}N_{T}}^{2} \simeq \sigma_{N_{d}} \sigma_{N_{T}}.
      \end{equation}

      \noindent Figure \ref{fig:Detections}b plots the detection fraction as a function of the stellar mass of our galaxies.  There is a clear, significant and sharp drop off in the fraction of galaxies with detected H$\alpha$ emission with M$_{*}>10^{11}$\,\msol.  In the most massive bin there are a total of 2 galaxies with H$\alpha$ detections ($12\pm25$ per cent) and 14 with no detections.  The redshifts of these galaxies are shown in Figure \ref{fig:MassiveZ}, and the galaxies detected in H$\alpha$ are all at around $z \sim$ 1.  The distribution in mass of the non-detections is similar to the whole sample.  Although our sample is small, there are no galaxies with M$_{*}>10^{11}$\,\msol that have H$\alpha$ detected at a redshift below $z = 0.93$ (Figure \ref{fig:MassiveZ}).
      
      Our H$\alpha$ detection fractions are plotted as a function of $(U-B)$ colour in Figure \ref{fig:Detections}c.  Bluer galaxies have a higher fraction with detected H$\alpha$ than redder ones.  It is clear that the majority of the galaxies detected in H$\alpha$ have a $(U-B) < 1.0$ and that the fraction of detections increases in bluer galaxies.  In the bluest bin there are more galaxies with detected H$\alpha$ than without, but for $(U-B) > 1.0$, there are more non-detections than detections.
      
      \begin{figure*}
        \centering
        \includegraphics[angle=0]{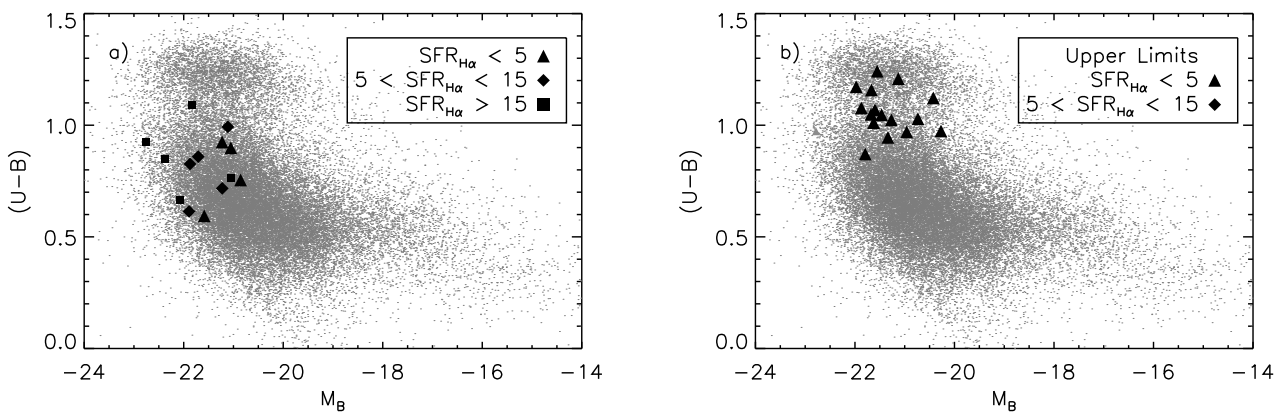}
        
        \caption{$(U-B)$ -- M$_{B}$ colour-magnitude diagrams for galaxies detected in H$\alpha$ (a), and for those not detected (b).  The gray points are the entire POWIR survey.  a) Detections -- triangle points have SFR$_{H\alpha,uncorr}$ $<$ 5\,\msol\,yr$^{-1}$, diamond points have SFR$_{H\alpha,uncorr}$ of 5-15\,\msol\,yr$^{-1}$ and square points have SFR$_{H\alpha,uncorr}$ $>$ 15\,\msol\,yr$^{-1}$.  b) Non Detections -- triangle points have SFR$_{H\alpha,uncorr}$ upper limit $<$ 5\,\msol\,yr$^{-1}$ and diamond points have SFR$_{H\alpha,uncorr}$ upper limit of 5-15\,\msol\,yr$^{-1}$.  SFRs are not corrected for dust.  There are no H$\alpha$ detections in the red sequence of galaxies.}
        \label{fig:colours}
      \end{figure*}
      
      Previously, Conselice et al. (2007, C07) investigated where on the $(U-B)$ verses $M_{B}$ colour-magnitude diagram galaxies in the POWIR survey fall as a function of mass.  They find evidence that galaxies evolve onto the red-sequence at $z <$ 1, but that galaxies with M$_{*}>10^{11.5}$\,\msol are generally always red, with a $\sim$40 per cent fraction that are blue at redshifts greater than $z \sim1.3$.
      
      Figure \ref{fig:colours} shows galaxies in this work over-laid on the colour-magnitude diagram for the entire POWIR survey, divided into those with detected H$\alpha$ emission, and those with H$\alpha$ upper limits (no detection).  Of the eight galaxies that were targeted and have continuum detections that fall within the red sequence, only one has detected star formation in H$\alpha$.  The majority of galaxies without H$\alpha$ detections lie in the red cloud or green valley.  Dust will move galaxies to redder colours as it attenuates the blue end of the spectrum more and so will complicate this plot.

    \subsection{Stacked spectra}
      \begin{figure}
        \centering
        \includegraphics[angle=0, width=8.5cm]{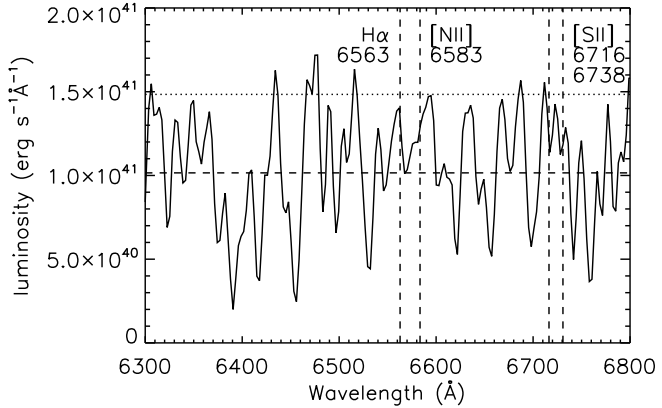}
        \caption{The averaged luminosity spectra for all galaxies with no H$\alpha$ detected.}
        \label{fig:StacksDetect}
      \end{figure}
      \begin{figure}
        \centering
        \includegraphics[angle=0, width=8.5cm]{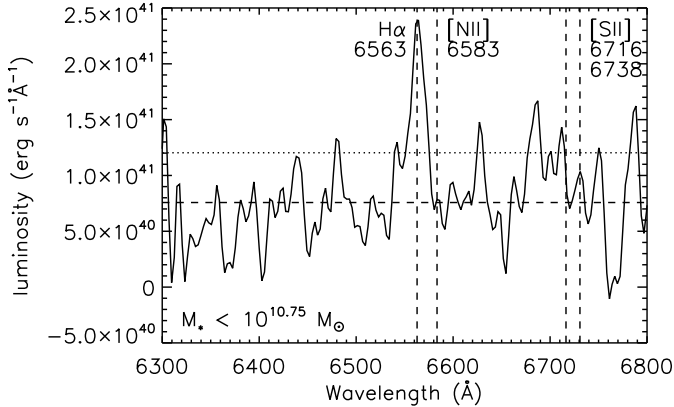}
        \caption{The averaged luminosity spectra for the lowest stellar mass bin, M$_{*}<10^{10.75}$\,\msol, of our sample.}
        \label{fig:StacksMass}
      \end{figure}
      \begin{figure}
        \centering
        \includegraphics[angle=0, width=8.5cm]{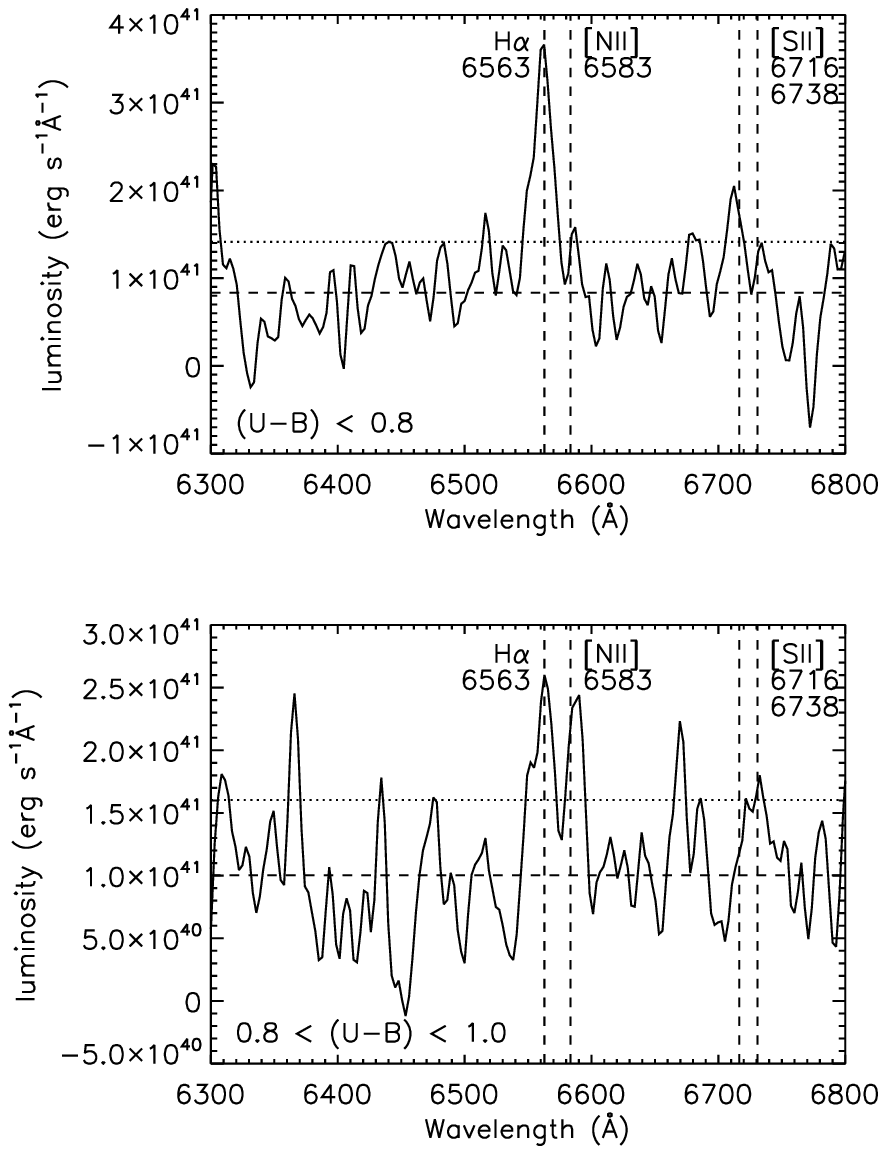}
        \caption{The averaged luminosity spectra for the bluest two colour bins in $(U-B)$ for the whole sample, $(U-B)<0.8$ (top) and $0.8\leq(U-B)<1.0$ (bottom).}
        \label{fig:StacksCol}
      \end{figure}

      To further investigate the galaxies where H$\alpha$ emission is not detected, the spectra from the galaxies without detected H$\alpha$ are stacked (Figure \ref{fig:StacksDetect}).  There is no detected H$\alpha$ emission in this averaged spectrum above the noise which equates to a mean SFR of $6.6$\,\msol\,yr$^{-1}$ (not dust corrected).  This value is lower than some individual galaxies with detected H$\alpha$ emission.  The non-detections have a mean redshift of 0.85 while the detections with a SFR less than 6.6 all have a redshift less than 0.81.  The mean redshift of the galaxies with detected H$\alpha$ emission is 0.78 and this difference in mean redshift is partially why the mean SFR limit for the non-detections is higher than the SFR in some of the galaxies with detected H$\alpha$ emission.
      
      Figure \ref{fig:StacksMass} shows the stacked spectra of the lowest stellar mass bin when the whole sample (detections and non-detections, not dust corrected) is stacked in stellar mass bins.  There is no significant change in H$\alpha$ emission in different stellar mass bins; these emission lines correspond to mean SFRs of $9.4 \pm 3.3$\,\msol\,yr$^{-1}$, $8.0 \pm 3.9$\,\msol\,yr$^{-1}$, and $12.1 \pm 5.3$\,\msol\,yr$^{-1}$ in the M$_{*}<10^{10.75}$\,\msol, $10^{10.75}\leq\mbox{M}_{*}/\mbox{\msol}<10^{11.0}$ and M$_{*}>10^{11.0}$\,\msol mass bins.
      
      Figure \ref{fig:StacksCol} shows the stacked spectra (bot dust corrected) for the bluest two bins when all the galaxies stacked in $(U-B)$ colour bins.  There is a clear drop in emission towards the redder bins corresponding to mean SFRs of $19.9 \pm 3.5$\,\msol\,yr$^{-1}$, $14.0 \pm 4.2$\,\msol\,yr$^{-1}$, $<7.3$\,\msol\,yr$^{-1}$ and $<6.3$\,\msol\,yr$^{-1}$ in the $(U-B)<0.8$, $0.8\leq(U-B)<1.0$, $1.0\leq(U-B)<1.2$ and $(U-B)>1.2$ colour bins.

    \subsection{Dust corrected star formation rates}

      \begin{figure}
        \centering
        \includegraphics[angle=0, width=8.5cm]{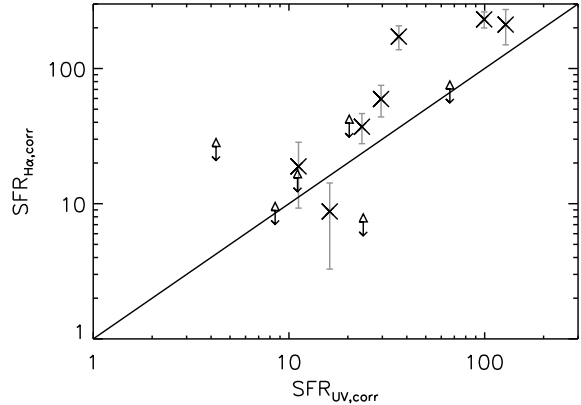}
        \caption{Comparison of the dust corrected H$\alpha$-derived SFRs with the dust corrected UV-derived SFRs for galaxies with UV measurements.  The triangular points with arrows are upper limits for the SFR$_{\rm{H\alpha,corr}}$.  There is good agreement between the two different measures of SFR.}
        \label{fig:ha_UV}
      \end{figure}

      \begin{figure*}
        \centering
        \includegraphics[angle=0]{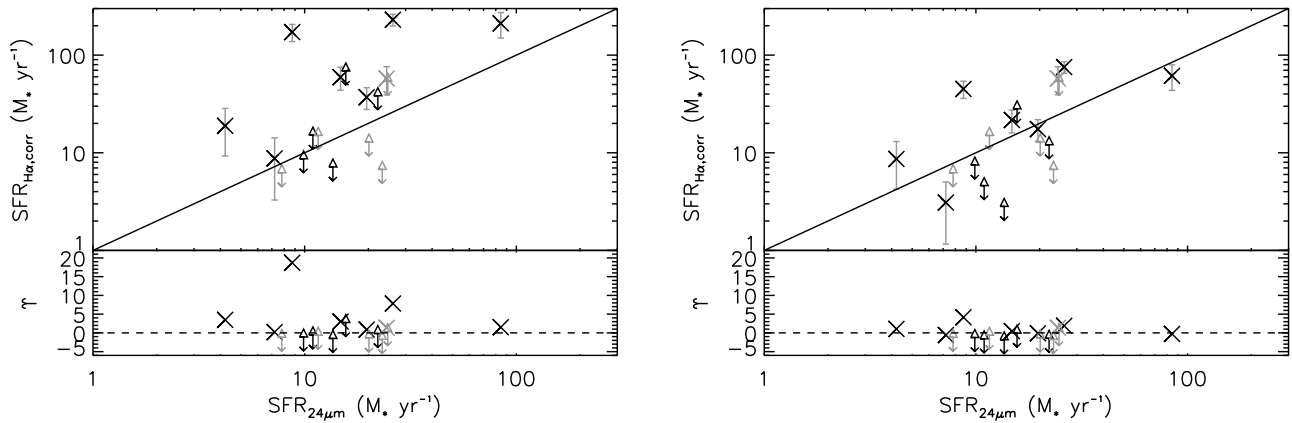}
        
        \caption{a) Comparison of the dust corrected H$\alpha$-derived SFRs with the 24\,$\mu$m-derived SFRs for galaxies with 24\,$\mu$m measurements.  Where the extinction correction has been derived from the $E(B-V)$-SFR relation (as opposed to calculating a galaxy's exact correction), the points are plotted in grey.  Underneath is plotted the SFR$_{H\alpha}$ excess $\Upsilon$, showing 2 galaxies with more star formation derived from H$\alpha$ than 24\,$\mu$m measurements.  b) shows the same but without the $E(B-V)_{stellar} = 0.44\,E(B-V)_{gas}$ correction being applied}
        \label{fig:ha_IR}
      \end{figure*}

      Figure \ref{fig:ha_UV} shows the comparison between the dust corrected H$\alpha$-derived and dust corrected UV-derived SFRs for the galaxies with UV measurements.  There is good agreement between the two SFR tracers.  Technically, for each galaxy, both SFR tracers have been dust corrected with the same $E(B-V)$ value, however, due to the dust attenuation curve being used, the actual amount both tracers have been adjusted by is different.  As there is a good agreement between the SFR measures, this helps confirm the accuracy of both measurements (see Bauer 2011).
      
      Figure \ref{fig:ha_IR}a shows the dust corrected H$\alpha$-derived SFRs plotted against the 24\,$\mu$m derived SFRs for the galaxies with available 24-$\mu$m measurements.  It is expected that the H$\alpha$ luminosity from star formation correlates with the 24-$\mu$m luminosity produced as a result of star formation.  However, other effects which also produce 24-$\mu$m emission would lead to an over prediction of the 24\,$\mu$m-derived SFR, such as an AGN, which however we do not find.
      
      Figure \ref{fig:ha_IR}a shows a relatively good correlation between the two measures of SFR apart from two points that lie with significantly larger SFR$_{H\alpha,corr}$ than SFR$_{24\,\mu m}$.  We quantify this difference by the parameter $\Upsilon = (\mbox{SFR}_{H\alpha,corr}-\mbox{SFR}_{24\,\mu m})/\mbox{SFR}_{24\,\mu m}$. It is not easy to see how points with a high (positive) $\Upsilon$ are physically possible and they are probably due to an overly large dust correction.
      
      
      It has also been suggested in some high redshift studies that the correction to the colour excess for stellar light (Equation \ref{eqn:E_BVstellar}) does not hold at high redshift, as it over-predicts H$\alpha$-derived SFRs with respect to the UV-derived SFRs (e.g. Erb et al. 2006, Hayashi et al. 2009, F\"{o}rster Schreiber et al. 2009).  The correction comes from the observation that the Balmer line emission is more attenuated (by a factor of $\sim2$) than starlight (see e.g. Petrosian, Silk \& Field 1972, Spitzer 1978, McKee \& Williams 1997).  When models are compared to observations, this effect appears to be due to the geometrical distribution of the attenuating dust in relation to the ionising stars (Calzetti et al. 1994, Charlot \& Fall 2000).  The fact that it may not hold at higher redshifts could be due to different dust geometries, or that galaxies at $z \sim2$ have a more top-heavy initial mass function than the IMF used (see for example Hayashi et al. 2009).
      
      Figure \ref{fig:ha_IR}b plots the H$\alpha$-derived SFRs when we do not apply this correction (i.e. when using $E(B-V)_{stellar}=E(B-V)_{gas}$) against the 24-$\mu$m derived SFRs.  These SFR$_{H\alpha,corr}$ values are lower and the correlation is now closer to what would be expected if 24-$\mu$m emission comes from both star formation and other sources.  This is similar to what is found in Erb et al. (2006) and Hayashi et al. (2009).  However Hayashi et al. conclude there is no firm evidence that $E(B-V)_{stellar} = 0.44\,E(B-V)_{gas}$ is incorrect.  It is not yet possible to measure the spatial dust content for galaxies at high redshift and thus there is no obvious way to validate any theory that the dust geometry in $z \sim$ 1 galaxies is different than in local galaxies.  We therefore use the 0.44 correction to the $E(B-V)_{gas}$ values in Equation \ref{eqn:E_BVstellar} for the remainder of this paper.

    \subsection{AGN heating and IR emission}
    
      Galaxies with an AGN are believed to heat dust close to the engine to a few thousand degrees (see for example Alonso-Herrero, 2001) which leads to an increase in the 24 $\mu$m IR emission.  Therefore, dusty AGN would be emitters of 24-$\mu$m radiation that could mimic highly star-forming galaxies.
      X-ray emission is used to distinguish AGN from normal galaxies in the EGS field due to both the emission from the relativistic jets and from accretion onto the central black hole.  While X-ray emission is a clear indicator of the presence of an AGN, the other types of emission can be attenuated significantly by dust.  This could lead to a large population of dusty AGN not detected in X-rays (see for example Wang \& Jiang, 2006).  At higher redshift, the lower received fluxes from these AGN could lead to more AGN being miss-classified as star-forming galaxies than at lower redshifts.
      
      To tell whether a galaxy contains an obscured AGN, the galaxy's 24-$\mu$m emission must be compared to a SFR tracer that is not strongly affected by the presence of an AGN.  H$\alpha$ is relatively insensitive to temperature and density of gas (Osterbrock 1989) and so is only slightly affected by the presence of an AGN (e.g. Dong et al. 2008) and thus a better representation of the SFR of the galaxy.  Figure \ref{fig:ha_IR}a shows the comparison of the 24-$\mu$m emission and dust corrected H$\alpha$ SFRs and SFR upper limits for the 18 galaxies that have an IR measurement, from the EGS field.  As the 24-$\mu$m emission can be caused by star formation, old stars or the presence of an AGN, it is expected that most galaxies with an AGN would have a higher 24\,$\mu$m-derived SFR than a dust corrected H$\alpha$-derived SFR.  There are three galaxies which have H$\alpha$-derived SFR upper limits that lie below the one-to-one relation line and could possibly have a 24-$\mu$m excess.  
      
      There are four galaxies that appear to have SFR$_{H\alpha}$ higher than SFR$_{24\,\mu m}$, two of these also have surprisingly high H$\alpha$/H$\beta$-derived $E(B-V)$ values.  These large Balmer decrement-derived values may be due to a larger than expected [NII]/H$\alpha$ ratio.   By assuming the UV-derived value of $E(B-V)$ is correct and comparing it with the Balmer decrement-derived value, a [NII]/H$\alpha$ ratio that would be required to reconcile the results can be obtained.  These values are shown in Table \ref{tab:nii/ha} for all the galaxies with available H$\beta$ data.
      
      \begin{table}
        \begin{center}
          \begin{tabular}{c c}
            \hline
            ID & $\log_{10}\left(\frac{F_{[NII]}}{F_{H\alpha}}\right)$ \\ \hline
            12024309 & 0.48 $\pm$ 0.32 \\
            12024440 & -0.61 $\pm$ 0.09 \\
            12024445 & 1.2 $\pm$ 1.8 \\
            12024528 & -0.90 $\pm$ 0.15 \\
            13035323 & $<$-0.50 \\ \hline
          \end{tabular}
        \end{center}
        \caption{[NII]/H$\alpha$ ratio for galaxies with H$\beta$ detection in the DEEP2 spectra.  Galaxy 13035323 only has an upper limit.}
        \label{tab:nii/ha}
      \end{table}
      
      The standard use of this type of line ratio diagnostic is carried out when plotted against the [OIII]/H$\beta$ ratio, for example by Kauffmann et al. (2003b) to separate AGN from normal galaxies.  The two galaxies that have a positive $\log(\mbox{[NII]/H$\alpha$})$ value would almost certainly be AGN using this classification; the three other galaxies appear to have $\log(\mbox{[NII]/H$\alpha$}) < -0.5$ and so probably fall into the star-forming region of the diagram.
      
      The Balmer decrement-derived $E(B-V)$ values are subject to large uncertainties due to the modest signal-to-noise ratio for the DEEP2 measured emission lines at this redshift.  The AGN fraction in the universe at these redshifts has been measured to be a few per cent (Silverman et al. 2009; Haggard et al. 2010) and so the number in a sample of star forming galaxies that have been solely selected by stellar mass will statistically be low.  It is still possible that they are highly obscured AGN but there is no evidence from our data that this may be the case, and thus we conclude that the AGN contamination of our sample is very small.

    \subsection{Star formation rate relations}
      \label{sec:StarFormation}
      
      \begin{figure*}
          \centering
          \includegraphics[angle=0]{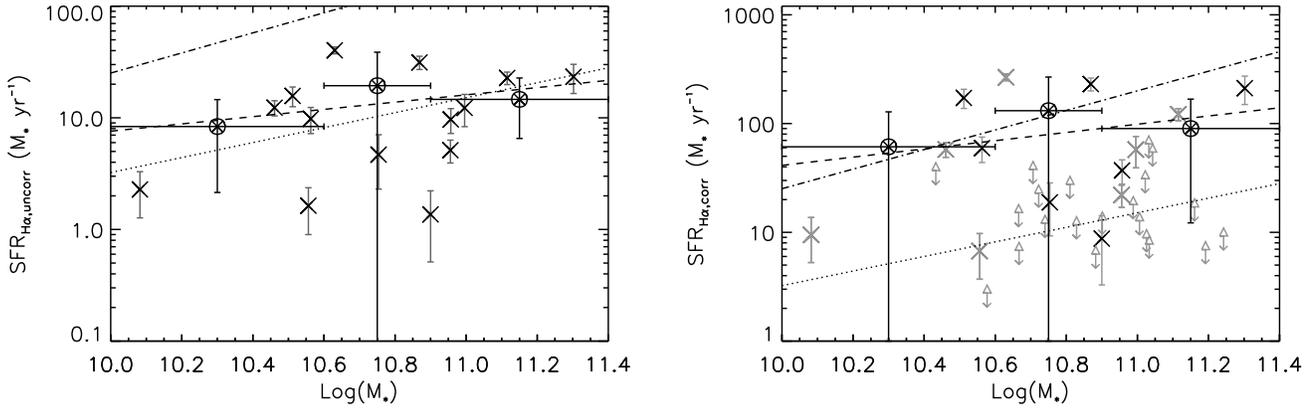}
          
          \caption{The uncorrected (a) and corrected (b) H$\alpha$ SFR plotted against stellar mass for all galaxies.  Galaxies with H$\alpha$ detections are plotted as crosses, the galaxies with an H$\alpha$ upper limit are plotted as a triangle with a downward pointing arrow.  Where the extinction correction has been derived from the $E(B-V)$-SFR relation (as opposed to calculating a galaxies exact correction), the points are plotted in grey (plot (b) only).  The large circled stars are the mean of the detections in mass bins and the best fit line for the galaxies with detections is plotted as the dashed line on each plot. The positions of the galactic main sequence as measured by Noeske et al. 2007a (N07, dotted) and Daddi et al. 2007 (D07, dot-dashed) are shown, as well as the best fit line for our galaxies with H$\alpha$ detections (dashed).}
          \label{fig:SFR_M}
      \end{figure*}
      
      Figure \ref{fig:SFR_M} shows the uncorrected and corrected H$\alpha$ SFRs as a function of stellar mass.  In both cases there is a possible slight increase of star formation with mass but is otherwise constant at all masses at M$_{*}>10^{10}$\,\msol.  The average points only use the measured SFRs, and do not include the upper limits, and these averages stay roughly constant with stellar mass.
      
      Previous observations have suggested that there is a ``Galactic Main Sequence'' (GMS) of star forming galaxies on a SFR - M$_{*}$ diagram (Noeske et al. 2007a hereafter N07, Daddi et al. 2007a hereafter D07, Pannella et al. 2009) and two of these lines are overplotted.  The N07 line (dots) was measured for a sample of galaxies in the redshift range $0.2< z <0.7$, and has a slope of $\log\left(\mbox{SFR}\right)\propto(0.67 \pm 0.08)\log\mbox{M}_{*}$.  They suggest that at redshift one, their measured GMS would appear to have the same slope but lie $\sim3$ times higher than their line for $0.2< z <0.7$.  Their GMS appears to lie below the majority of our galaxies, suggesting that our galaxies have higher measured SFR at a given stellar mass.
      
      The line from D07 (dashed) has a slope of $\log\left(\mbox{SFR}\right)\propto0.9\log\mbox{M}_{*}$ which is similar to F\"{o}rster Schreiber et al (2009) who found it to be close to their H$\alpha$ derived values.  The line lies close to our dust-corrected detections, however it appears to have a steeper slope than our galaxies.
      
      We use a Monte Carlo simulation to determine the probability that our data fits the N07 GMS.  Each data point is moved randomly within its error bar and a line of best fit is found using a least-squares method.  This is repeated $10^{4}$ times and the mean slope and intercept point are calculated as:
      
      \begin{equation}
        \log\left(\mbox{SFR}_{H\alpha,corr}\right)=\left(0.40\pm 0.13\right)\log\mbox{M}_{*}-\left(2.4\pm 1.3\right), 
      \end{equation}
      
      \noindent suggesting a slope slightly flatter than found in previous work.  There is a 8 per cent probability that the slope for our data is within the errors of the slope of N07.  There is no confidence greater than the 1 $\sigma$ level that our derived slope is different to the slope of N07, so we cannot confirm that the two are the same.
      
      \begin{figure*}
          \centering
          \includegraphics[angle=0]{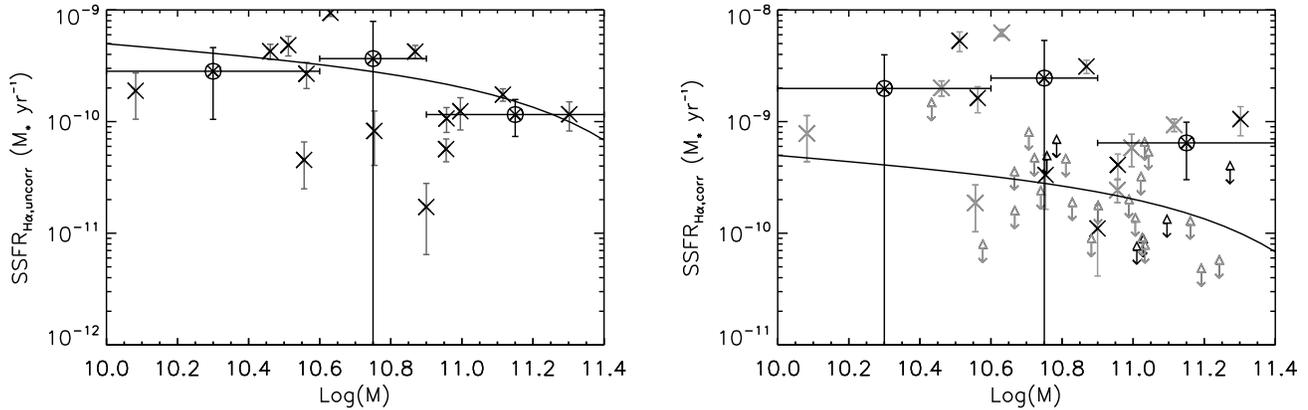}
          
          \caption{The uncorrected (a) and corrected (b) H$\alpha$ SSFR plotted against stellar mass for all galaxies.  The points are the same as in Figure \ref{fig:SFR_M}.  The solid line is the modeled relation for galaxies with an exponential star formation history, as discussed in Noeske et al. (2007b).}
          \label{fig:SSFR_M}
      \end{figure*}
      
      A measure of the formation histories of galaxies can be inferred from the specific star formation rate (SFR divided by stellar mass, SSFR) which is plotted against mass in Figure \ref{fig:SSFR_M}.  The SSFR takes into account the fact that the same SFR will have a different effect on galaxies with different stellar masses and so gives a clearer picture of the SFR history of the galaxies probed.
      
      The SSFR appears to fall with increasing mass, especially at M$_{*}>10^{10.9}$\,\msol, and the mean value of the detections is possibly constant at masses lower than this.  Along with the previous finding that there are no galaxies with M$_{*}>10^{11}$\,\msol that have H$\alpha$ detected at a redshift below $z=$ 0.93 in the sample, this suggests that at $z \sim$ 1, galaxies with M$_{*}\gtrsim10^{11}$\,\msol are much less likely to still be forming stars; but where they are, they are forming stars at a similar relative rate to lower mass galaxies.
      
      Overplotted on Figure \ref{fig:SSFR_M}b is the theoretical relation from the exponential star formation history model of Noeske et al. (2007b) at a redshift of $z = 0.8$.  This model uses a mass dependent formation redshift for galaxies and evolves their star formation using a closed-box model.  The model appears to fit the relation observed in our data.  However, the dust corrected SSFRs for galaxies with H$\alpha$ detections are possibly slightly higher.  When taking into consideration the upper limit points, the model is a slightly better fit.  Our points do though lie within the scatter of the data used to derive this relation in Noeske et al.
      
      \begin{figure*}
          \centering
          \includegraphics[angle=0]{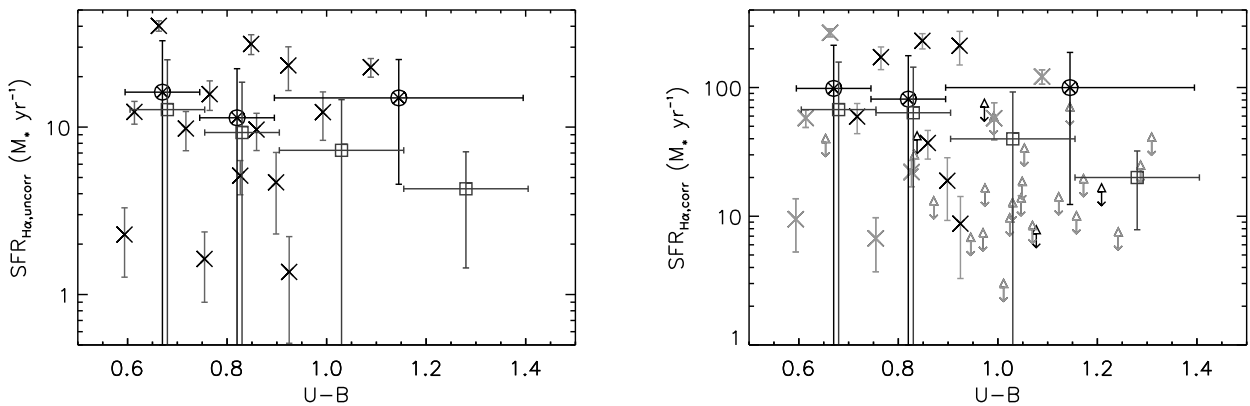}
          
          \caption{The uncorrected (a) and corrected (b) H$\alpha$ SFR plotted against colour for all galaxies.  The points are the same as in Figure \ref{fig:SFR_M}.  The large circled stars are the mean of the H$\alpha$ detections in mass bins, the square points are the mean SSFR of all galaxies in mass bins, including the upper limits -- See text for details.  The circled stars and square points are shifted by $\pm$0.005 in the x axis for clarity.}
          \label{fig:SFR_colour}
      \end{figure*}
      
      In Figure \ref{fig:SFR_colour}, the SFR is plotted against $(U-B)$ colour along with the mean value of the detections.  Also plotted is the mean colour of all the points.  As the number of H$\alpha$ detections drop with increasing $(U-B)$ colour (Figure \ref{fig:Detections}c), the number of galaxies with H$\alpha$ detections is very low at $(U-B)\gtrsim0.9$.  The galaxies that are detected are the galaxies with the most on-going star formation and so are not necessarily representative of all galaxies.  Taking the mean of all the points (detections and upper limits) does not necessarily give a meaningful number but it better shows the overall trend alluded to by the data points.  It also allows the large $0.9\leq (U-B) \leq1.4$ bin to be split into two.  It is clear that the two methods show different trends with the mean value of detections constant and the mean value of all points dropping for redder galaxies.
      
      Comparing the two methods, there are apparently two populations of galaxies, those that are star forming which have roughly the same SFR whatever their colour; and galaxies that are not star forming.  At redder colours, there are more galaxies not star forming which lowers the average SFR of all the galaxies at very red $(U-B)$.  
      
      \begin{figure*}
          \centering
          \includegraphics[angle=0]{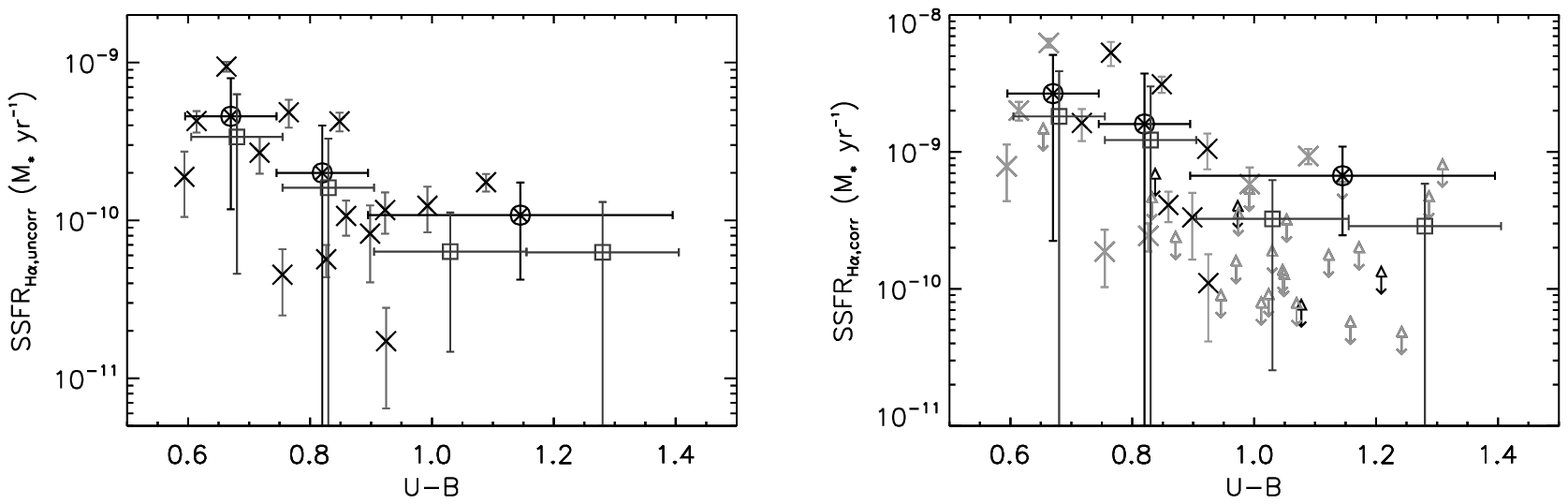}
          
          \caption{The uncorrected (a) and corrected (b) H$\alpha$ SSFR plotted against $(U-B)$ colour for all galaxies.  The points are the same as in Figure \ref{fig:SFR_colour}.}
          \label{fig:SSFR_colour}
      \end{figure*}
      
      To look at these effects more productively, the SSFR is compared to the $(U-B)$ colour in Figure \ref{fig:SSFR_colour}.  The SSFR is clearly lower in redder galaxies and the trend is similar for both the mean value of detections, and the mean value of all points.

    \subsection{CAS morphology}
      
      \begin{figure*}
        \centering
        \includegraphics[angle=0]{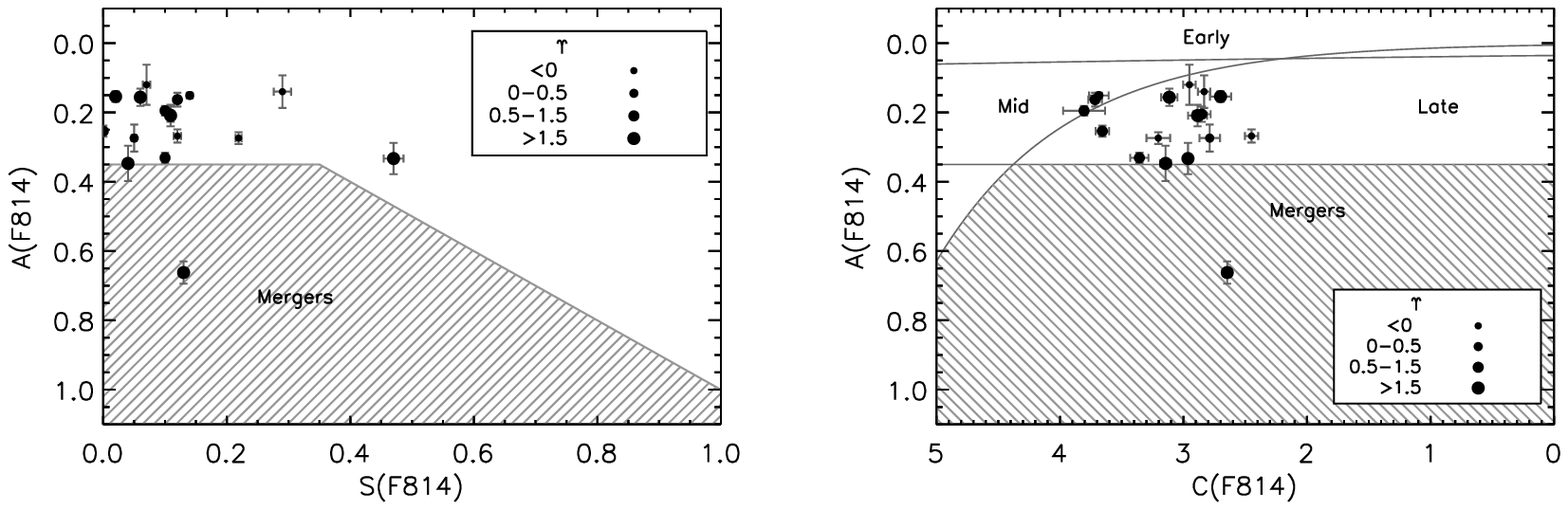}
        
        \caption{a) CAS asymmetry against clumpiness parameters for the galaxies with 24-$\mu$m measurements.  The parameter space for mergers is defined as A $>0.35$ and A $>$ S. The points are sized by the SFR excess $\Upsilon$.  b) Concentration against asymmetry for the same galaxies.  Mergers are defined as A $>0.35$.  The boundary between Early and Mid/Late type galaxies is $\mbox{C}=21.5\log_{10}(\mbox{A})+31.2$ (Conselice et al. 2008b).}
        \label{fig:CAS}
      \end{figure*}
      
      \begin{figure*}
        \centering
        \includegraphics[angle=0]{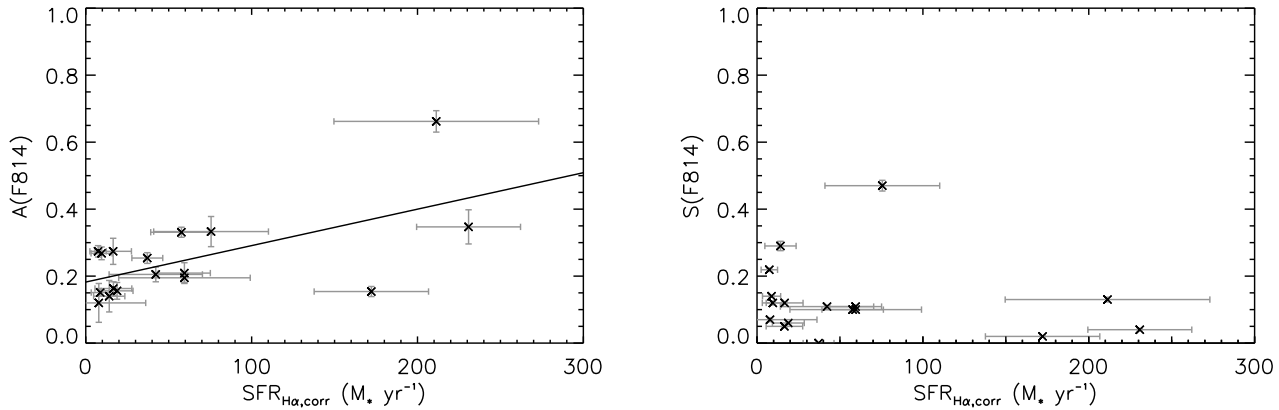}
        
        \caption{a) CAS Asymmetry against corrected SFR for the same galaxies as Figure \ref{fig:CAS} with the best fit line shown.  b) CAS Clumpiness against corrected SFR for the same galaxies.}
        \label{fig:CAS_SFR}
      \end{figure*}
      
      The CAS morphology parameters (Conselice 2003) are analysed on our sample using the \textit{ACS} imaging to determine whether any of our sample are likely mergers which may affect the SFR determinations.  These parameters also allow us to examine the nature of the star formation we measure and investigate the morphologies of the galaxies.
      
      Within the asymmetry (A) against clumpiness (S) plot, a galaxy merger is defined as having A $>$ 0.35 and S $>$ A (Conselice, Rajgor \& Myers 2008b).  For a concentration (C) against asymmetry plot, a galaxy merger is defined as having A $>$ 0.35 and C $<2.44\log(\mbox{A})+5.49$. (Conselice et al. 2008b).  These plots are shown in Figure \ref{fig:CAS} along with the position of the three galaxies identified from the imaging; the points are coloured by bins of SFR$_{H\alpha}$ excess $\Upsilon$.
      
      Plotted on Figure \ref{fig:CAS}b is the line dividing early- and mid/late-type galaxies as defined in Conselice et al. (2008b).  It is clear that all of the galaxies we observe that are not probable mergers, are mid- or late-type spirals which is confirmed by the visual morphology (Figure \ref{fig:Images}).
      
      There is no clear relation between the SFR$_{H\alpha}$ excess $\Upsilon$ and the position in the CAS parameter space.  The SFR excesses are therefore likely due to the individual circumstances of each galaxy.
      
      The possible merging galaxy (12024436) is confirmed to be a merger by the CAS parameters.
      Two other galaxies lie very close to the boundary of the merger parameter space - galaxies 12024445 where the upper error bar extend into the merger region, and 13035323.  Galaxy 13035323 does not have an unusual SFR$_{H\alpha,corr}$ however galaxy 12024445 is one of the galaxies with a larger SFR$_{H\alpha,corr}$ than SFR$_{24\,\mu m}$.
      
      The possible merging systems 12020031 and 13026831 are not mergers as found in CAS parameter space.
      The double galaxy system 12020027 does not appear to be a merging system likely because the CAS code ignored the other galaxy (see Conselice 2006 for time-scale issues within CAS).
      
      Figure \ref{fig:CAS_SFR} shows the asymmetry and clumpiness of our galaxies as a function of their measured SFRs.  There is possibly a relation between the asymmetry and the SFR and the galaxies are fit by a line with the relation
      
      \begin{equation}
        \mbox{A}=\left(1.09\pm0.36\right) \times 10^{-3}\, \mbox{SFR}_{\rm{H\alpha,corr}} + \left(0.182\pm0.034\right)
      \end{equation}
      
      \noindent This is likely to be due to the star formation happening in an asymmetric manner -- the more star formation happening, the more asymmetric the galaxy is -- or that what ever mechanism is causing the asymmetry is driving the star formation.  If this were the case, it would point to star formation not being `triggered' by some (possibly small) event that then causes a large 'snowballing' of star formation activity, but a regime where larger events lead to larger star formation bursts.  In reality both effects probably happen accentuating the effect: star-formation triggering events cause the galaxy to look more asymmetric and trigger asymmetric star-formation episodes.
      
      Somewhat surprisingly then there is no obvious relation between a galaxy's SFR and its clumpiness (Figure \ref{fig:CAS_SFR}b).  This might imply that star formation is happening in large, less well defined regions of the galaxy.  There is no  relation observed between the concentration and SFR.

      \begin{table*}
        \begin{center}
          \begin{tabular}{c c c c c c c p{4.5cm} }
            \hline
            ID & $E(B-V)$ & SFR$_{H\alpha,corr}$ & $\Upsilon$ & C & A & S & Notes \\ \hline
            12020027 & 1.02 & 101 $\pm$ 31 & 10.6 & 2.70 $\pm$ 0.09 & 0.15 $\pm$ 0.02 & 0.02 $\pm$ 0.002 & Two galaxies in image\\
            12020031 & 0.68 & 44 $\pm$ 23 & 1.84 & 2.96 $\pm$ 0.03 & 0.33 $\pm$ 0.05 & 0.47 $\pm$ 0.02 & Possible second galaxy in image.\\
            12024436 & 0.94 & 124 $\pm$ 55 & 0.472 & 2.65 $\pm$ 0.03 & 0.66 $\pm$ 0.03 & 0.13 $\pm$ 0.01 & Possible star in slit.  CAS merger.\\
            12024440 & 0.60 & 11.1 $\pm$ 6.4 & 1.63 & 3.12 $\pm$ 0.07 & 0.16 $\pm$ 0.03 & 0.06 $\pm$ 0.003 & \\
            12024445 & 0.85 & 136 $\pm$ 30 & 4.20 & 3.15 $\pm$ 0.04 & 0.35 $\pm$ 0.05 & 0.04 $\pm$ 0.004 & Possible CAS merger.\\
            13026831 & 0.66 & 52 $\pm$ 37 & 1.10 & 3.81 $\pm$ 0.17 & 0.20 $\pm$ 0.01 & 0.10 $\pm$ 0.002 & Poss merger in image, not in CAS.\\
            13035323 & 0.64 & 32 $\pm$ 19 & 0.306 & 3.38 $\pm$ 0.07 & 0.33 $\pm$ 0.02 & 0.10 $\pm$ 0.004 & Possible CAS merger.\\ \hline
          \end{tabular}
        \end{center}
        \caption{Galaxies with high H$\alpha$-derived SFR excess, $\Upsilon$, have morphological features, or identified as mergers by CAS parameters.}
        \label{tab:SFRexcess}
      \end{table*}

  \section{Discussion}
    \label{sec:Discussion}

    \subsection{The shut-off of star formation in galaxies}

      In the scenario of downsizing  the most massive galaxies should turn off their star formation at the redshift ranges we probe.  In our sample, the most massive galaxies with star formation and with stellar masses M$_{*}>10^{11}$\,\msol are found only at a redshift greater than $z$ = 0.93.  The drop-off in the number of star forming galaxies with masses above M$_{*} = 10^{11}$\,\msol is sharp, suggesting that the process of turning off star formation happens very quickly, in less than a Gyr.  This appears to be a sudden quenching rather than a slow ending of star formation, similarly it is not likely to be a ``running out'' of gas within star forming regions as in this scenario we would expect this to lead to a declining SFR over a longer period of time. 
      
      There is only one galaxy that is detected in H$\alpha$ and lies within the red sequence of galaxies in our sample, showing that star forming galaxies detected with H$\alpha$ at z$\sim$1 are mostly blue.  There are only a small number of non-detections (4 out of 25) that lie definitively within the region of blue galaxies and these could be either non-detections of star forming galaxies or actual non-star forming galaxies. 

      In line with what other studies have found at other wavelengths, this study finds a slightly increasing H$\alpha$-derived SFR with stellar mass, and a decreasing SSFR with mass relations.  There is some suggestion that the normalisation of the `galactic main sequence' (GMS, N07) is dependent on the wavelength and method used to measure the SFR, and the H$\alpha$-derived SFRs appear to be slightly higher than other measurements.  This could be the result of the sensitivity of measuring SFRs which in this study are limited to greater than a few \msol\,yr$^{-1}$.  With a dust correction, this becomes a lower limit of SFRs greater than a few tens of \msol\,yr$^{-1}$.  [OII] star formation rates can be measured (uncorrected) typically down to 0.5--1\,\msol\,yr$^{-1}$, which when corrected give SFRs with typically half the value detected in this sample.  However, to reconcile our results with the N07 derived GMS, there would have to be a significant number of galaxies which have a SFR an order of magnitude below the level of our GMS at all masses.
      
      There are two probable causes of our detections having larger average SFRs than the GMS measured by N07.  A major contributor will be that some of our galaxies have been selected to have significant 24-$\mu$m emission, meaning that they are either highly star forming, or have a significant AGN component.  Two galaxies (12024309 \& 12024445) could have [NII]/H$\alpha$ ratios indicative of AGN, but have a SFR$_{H\alpha}$ excess suggesting otherwise.  These two therefore probably have a slightly larger amount of [NII] emission leading to a overestimation of the H$\alpha$-derived SFR.  There is little other evidence of any other galaxies in this sample having a dominating AGN and so the higher SFRs measured are likely because the galaxies actually have a higher rate of star formation.  These would then lie above (i.e. with a higher SFR) any GMS relationship found in more complete samples.  The galaxies however lie much above the N07 GMS (even with their estimated raising of the normalisation for higher redshifts), significantly beyond their 0.3 dex estimated scatter.
      
      The Daddi et al. (2007, D07) GMS is derived from UV and 24-$\mu$m emission(as opposed to [OII] and 24-$\mu$m in N07) and appears to be at a similar level to our data, but it is not clear whether our data follows their slope.  It is clear then that the choice of SFR indicator can be important.  C07 measured that the 24\,$\mu$m-derived SFR were considerably higher than the [OII] values and it is not unusual that the two methods do not give agreement on a galaxy-by-galaxy basis (e.g. Salim et al. 2007).
      
      The SSFR is a measure of how much of a galaxy's mass has been formed in recent star formation episodes.  The most massive galaxies with M$_{*} > 10^{11}$\,\msol have a lower SSFR than the less massive galaxies which have a roughly constant value.  These very massive galaxies have had less of their mass formed in recent star formation and so must contain older stellar populations. This is a clear indication that the most massive galaxies had their star formation quenched at a higher redshift than lower mass galaxies.

    \subsection{Colour transition}

\begin{figure}
        \centering
        \includegraphics[angle=0, width=8.5cm]{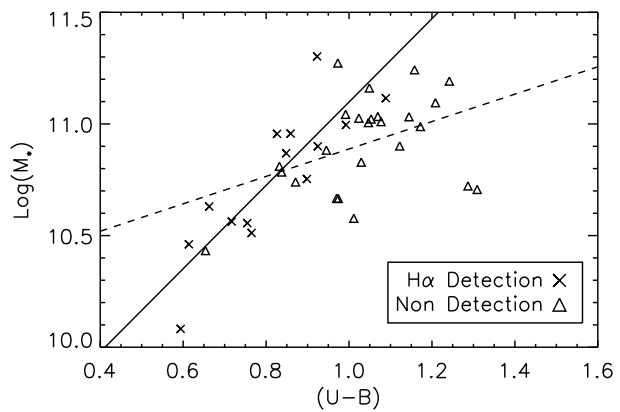}
        
        \caption{Stellar mass against $(U-B)$ colour for galaxies divided into those with detected star formation (crosses) and no detected star formation (triangles).  The redder galaxies tend to be more massive than the blue galaxies -- this is possibly a selection effect due to half our galaxies needing 24\,$\mu$m emission for selection and thus have star formation.  There is a lack of massive, blue galaxies which should have been detected.  The solid line is a least-squares fit to the galaxies with H$\alpha$ detections and has a slope of $\log \left(\mbox{M}_{*}\right) \propto \left( 1.87 \pm 0.33\right) \times (U-B)$.  The dashed line is a least-squares fit to the galaxies without detected H$\alpha$ and has a smaller slope: $\log \left(\mbox{M}_{*}\right) \propto \left( 0.91 \pm 0.18\right) \times (U-B)$}
        \label{fig:ColourMass}
      \end{figure}

      We find a separation between the star forming and non-star forming galaxies in SFR-colour plots (Figure \ref{fig:SFR_colour}).  Galaxies that are towards the red end of our sample (in $(U-B)$) and have measured star formation, have the same average SFR as the blue galaxies.  In the SSFR-colour plots (Figure \ref{fig:SSFR_colour}), the SSFR drops with colour due to the fact that the red star forming galaxies are more massive than the blue galaxies with the same SFR.  This is seen in Figure \ref{fig:ColourMass} and is possibly a selection effect.  Half the galaxies in the sample were selected to have a 24-$\mu$m detection -- for a galaxy to have enough star formation so that it gives 24-$\mu$m emission, but not be significant enough to make the galaxy blue, the galaxy must be more massive.  Figure \ref{fig:ColourMass} also shows a considerable difference in the slope of this relation for star-forming and passive galaxies.  The star-forming galaxies are fit by a line with the relation
      
      \begin{equation}
        \log \left(M_{*}\right) = (1.87 \pm 0.33) \times (U-B) + (9.23 \pm 0.28),
      \end{equation}
      
      \noindent and the galaxies without H$\alpha$ detections are fit by the relation
      
      \begin{equation}
        \log \left(M_{*}\right) = (0.61 \pm 0.27) \times (U-B) + (10.27 \pm 0.28).
      \end{equation}
      
      \noindent A two dimensional Kolmogorov--Smirnov test (K-S test; Fasano \& Franceschini, 1987) indicates that these two populations are however not significantly different.

      \begin{figure*}
        \centering
        \includegraphics[angle=0, width=17cm]{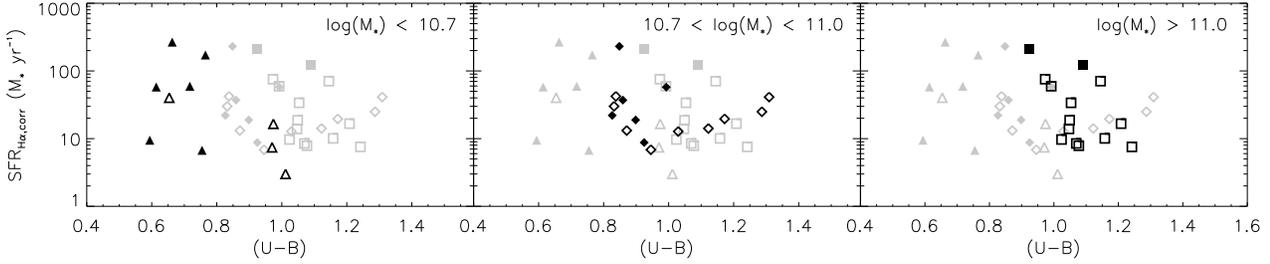}
        
        \caption{SFR vs colour binned by stellar mass.  Filled in symbols have detected H$\alpha$ emission, empty symbols do not.  The black points are the galaxies within the stellar mass range of the bin: Triangles -- M$_{*} \leq 10^{10.7}$\,\msol, diamonds -- $10^{10.7}\mbox{\,\msol} \leq \mbox{M}_{*} \leq 10^{11.0}$\,\msol, squares -- M$_{*} \geq 10^{11.0}$\,\msol; the light grey points are the galaxies in the other mass bins shown for comparison.  The lowest mass bin contains the highest fraction of star forming galaxies, the highest mass bin contains the highest fraction of passive galaxies.  Within each mass bin, galaxies that are star forming have similar $(U-B)$ colours and a range of SFRs; galaxies that have finished their star formation have a wide range of colours.}
        \label{fig:SFRcolourMass}
      \end{figure*}

      A more detailed look at the SFR and SSFR against colour relations can be obtained by binning the points by stellar mass (Figure \ref{fig:SFRcolourMass}).  The effects of downsizing are clearly seen in these plots with the highest mass bins having many more passive galaxies than star-forming ones, similarly the lowest mass bin has more star-forming than passive galaxies.  Both the lowest and intermediate mass bins in the SFR-colour plot can be described as a number of star-forming galaxies with a range of SFRs but with similar colours, with the passive galaxies (with similar SFRs) having a range of $(U-B)$ colours.  The most massive galaxies are mostly passive, red galaxies.
      
      When galaxies turn off their star formation, they also usually start changing colour from blue to red.  This would lead to the galaxies moving downwards and to the right in the SFR-colour diagram.  If the concept of downsizing is correct, then the most massive galaxies shut off their star formation before less massive systems.  We can then use a mass cut to separate a population of galaxies that have shut off their star formation (the most massive galaxies) from a population that are yet to shut off their star formation (the least massive galaxies).  Figure \ref{fig:SFRcolourMass} splits the galaxies into stellar mass bins and shows their position on a SFR-colour diagram.
      
      In each stellar mass bin, the star forming galaxies have a bluer $(U-B)$ colour than the passive galaxies and the range of colours that the star forming galaxies have in each stellar mass bin is quite narrow.  Passive galaxies have a range of $(U-B)$ colours.  Each stellar mass bin is successively redder, this is due to the relation between stellar mass and colour (Figure \ref{fig:ColourMass}).

      As seen in Figure \ref{fig:SSFR_colour} there is a relationship between the SSFR and colour.  This is partially expected -- SSFR traces the effect that star formation has had on the the recent history of the galaxy.  This will also be borne out in the colour of the galaxy, with the more recent the star formation, the more affect it has on a galaxy's colour.  This then explains why the different mass bins are separated along the colour axis of the SFR-colour diagram (Figure \ref{fig:SFRcolourMass}): the redder systems have lower SSFRs and so for the same SFR they must be more massive than bluer systems.
      
      This relation could be affected by our selection criteria.  We could be missing low-mass passive objects which are red.  However due to their low mass, they would not lie considerably below galaxies already plotted on the SSFR-colour diagram.
      
      The galaxies investigated appear to follow what would be expected in a downsizing senario.  There are significantly fewer galaxies with M$_{*}>10^{11}$\,\msol forming stars and the most massive galaxies are only forming stars at the highest redshifts probed.  More detailed studies into the SFR-M$_{*}$ as a function of redshift are not possible due to the small size of our sample.
    
%
  
  \section{Conclusions}
    \label{sec:Conclusion}

    We have analysed NIR spectroscopy of 41 massive (M$_{*}>10^{10}$\,\msol) galaxies taken from the POWIR Survey of which half were chosen to have significant 24-$\mu$m flux, no X-ray emission and within the range $0.4<\mbox{z}_{spec}<1.3$.  We have acquired NIR spectroscopic data for these systems and measured SFRs from the H$\alpha$ emission line.  The H$\alpha$ emission line is the best direct measure of instantaneous star formation as it is coupled directly to the massive stars in formation, is largely independent of metalicity and is less effected by dust attenuation than either the [OII] emission line or UV flux.
    
    We investigated several methods of correcting high-redshift galaxies for dust using multi-wavelength methods and chose to ultimately apply a method derived from UV data.  We find that the H$\alpha$ SFR measurements are in good agreement with those measured by UV emission, but are slightly higher than those measured from 24-$\mu$m emission.  By not using the correction to the $E(B-V)$ value due to differences associated with gas emission and stellar light (Calzetti et al. 2000) our H$\alpha$ corrected SFRs are similar to or less than the galaxies'  24-$\mu$m derived SFRs.  The AGN contamination of our sample is low -- although highly obscured AGN may exist within the sample there is no evidence that there are any.
    
    We find that there is a drop in the fraction of galaxies with detections of H$\alpha$ for systems with a stellar mass greater than M$_{*}=10^{11}$\,\msol.  The massive galaxies with detected H$\alpha$ are all found at redshifts greater than $z=$ 0.93.  The fraction of galaxies with detected H$\alpha$  also drops for galaxies that are red in U-B colour.  The sharpness of this drop implies that star formation is truncated over a very short period of time rather than dropping steadily over an extended epoch.
    
    We investigated the H$\alpha$ SFR as a function of stellar mass and find that SFR stays roughly constant with stellar mass.  Likewise, the SSFR drops with increasing mass, especially at M$_{*}>10^{10.9}$\,\msol.  This implies that at $z \sim$ 1, galaxies with M$_{*}\gtrsim10^{11}$\,\msol are less likely to be forming stars, and when they are, they are doing so at a much lower rate relative to their mass.  These massive galaxies have a lower SSFR than lower mass galaxies, implying that they have on average a  stellar population that formed earlier, and must have had their star formation quenched at a higher redshift than the lower mass galaxies.
    
    We find that the observed slope in the SFR-mass relation appears to be slightly flatter than what is found in previous studies.  The SSFR-mass relation also agrees with the relation predicted by an exponentially decreasing star formation history model with mass dependent formation redshifts.  Our measured SSFRs however are slightly higher than the Noeske et al. (2007b) model predictions.
    
    We also investigate the SFR as a function of colour and find that the galaxies that are detected in H$\alpha$ form stars at similar rates independent of  colour.  When split into bins of stellar mass, within each bin we find star forming galaxies that have a narrow range of $(U-B)$ colours while galaxies not detected in H$\alpha$  have a wider range of colours which are mostly redder than the star forming ones.  These mass bins can be viewed as populations of galaxies in different stages of their evolution from star forming to passive, and suggests that galaxies drop in star formation before becoming redder.   We also find that the specific SFR drops with increasing $(U-B)$ colour (redness) for all galaxies; this is probably due to both being a measure of the effect on the galaxy of recent star formation.  This relation leads to the more massive galaxies (with lower SSFR) becoming redder.  Evolutionary tracks on the SFR-colour plot are therefore shifted towards redder colours for the most massive galaxies.  We also investigate the structures of our sample which have ACS HST imaging.    We find that there is possible relation found between a galaxy's SFR and its asymmetry as measured by the CAS parameters (Conselice 2003).
    
    H$\alpha$-derived SFRs at $z \sim 1$ appear to be similar to those measured using other tracers.  There is a large body of work studying star formation at these redshifts using these other tracers, and we have shown, using the more reliable H$\alpha$ values  that these other results are consistent at least up to $z \sim 1.4$.  As NIR spectroscopic instrument technology continues to evolve, especially on 8-10m telescopes, the growth in the number and quality of H$\alpha$ measurements, particularly at high redshift, will allow us to compare these different SFR tracers at higher redshifts and therefore obtain an even more accurate measure of the star formation history of galaxies.
  
  \section*{ACKNOWLEDGMENTS}

  Funding for the trip to the Obseratorio del Roque de los Muchachos was provided by the University of Nottingham.  We thank Nathan Bourne for useful and informative discussions.  We acknowledge support from the STFC as well as a Leverhulme Trust Prize to CJC.  The William Herschel Telescope is operated on the island of La Palma by the Isaac Newton Group in the Spanish Observatorio del Roque de los Muchachos of the Instituto de Astrof\'{\i}sica de Canarias.




  \label{lastpage}


\begin{thebibliography}{99}
 
  \bibitem[\protect\citeauthoryear{}{}]{b1}   Acosta-Pulido J. A., Ballesteros E., Barreto M. et al., 2003, INGN, 7, 15
    
    \bibitem[\protect\citeauthoryear{}{}]{Afonso03}
      Afonso J., Hopkins A., Mobasher B., Almeida C., 2003, ApJ, 597, 269 
    
    \bibitem[\protect\citeauthoryear{}{}]{Alonso-Herrero01}
      Alonso-Herrero A., Quillen A. C., Simpson C., Efstathiou A., Ward M. J., 2001, AJ, 121, 1369
    
    \bibitem[\protect\citeauthoryear{}{}]{Baldwin81}
      Baldwin J. A., Phillips M. M., Terlevich R., 1981, PASP, 93, 5
    
    \bibitem[\protect\citeauthoryear{}{}]{Bauer11}
      Bauer A. E., Conselice C. J., Perez-Gonzalez P. G., Grutzbauch R., Bluck A. F. L., Buitrago F., Mortlock A., 2011, arXiv:1106.2656v1
    
    \bibitem[\protect\citeauthoryear{}{}]{Bell07}
      Bell E. F., Zheng X. Z., Papovich C., Borch A., Wolf C, Meisenheimer K., 2007, ApJ, 663, 834
    
    \bibitem[\protect\citeauthoryear{}{}]{Bevington69}
      Bevington P. R., 1969, Data Reduction and Error Analysis for the Physical Sciences. McGraw-Hill, New York
    
    \bibitem[\protect\citeauthoryear{}{}]{Bundy06}
      Bundy K., Ellis R. S., Conselice C. J. et al., 2006, ApJ, 651, 120
    
    \bibitem[\protect\citeauthoryear{}{}]{Calzetti01}
      Calzetti D., 2001, PASP, 113, 1449
    
    \bibitem[\protect\citeauthoryear{}{}]{Calzetti94}
      Calzetti D., Kinney A. L., Storchi-Bergmann T., 1994, ApJ, 429, 582
    
    \bibitem[\protect\citeauthoryear{}{}]{Calzetti00}
      Calzetti D., Armus L., Bohlin R. C., Kinney A. L., Koornneef J., Storchi-Bergmann T., 2000, ApJ, 533, 682
    
    \bibitem[\protect\citeauthoryear{}{}]{Calzetti07}
      Calzetti D., Kennicutt R. C., Engelbracht C. W. et al., 2007, ApJ, 666, 870
    
    \bibitem[\protect\citeauthoryear{}{}]{Cardelli89}
      Cardelli J. A., Clayton G. C., Mathis J,S., 1989, ApJ, 345, 245
    
    \bibitem[\protect\citeauthoryear{}{}]{Chabrier03}
      Chabrier G., 2003, PASP, 115, 763
    
    \bibitem[\protect\citeauthoryear{}{}]{Chapman05}
      Chapman S. C., Blain A. W., Smail I., Ivison R. J., 2005, ApJ, 622, 772
    
    \bibitem[\protect\citeauthoryear{}{}]{Charlot00}
      Charlot S., Fall S. M., 2000, ApJ, 539, 718
    
    \bibitem[\protect\citeauthoryear{}{}]{Charlot02}
      Charlot S., Kauffmann G., Longhetti M., Tresse L., White S. D. M., Maddox S. J., Fall S. M., 2002, MNRAS, 330, 876
    
    \bibitem[\protect\citeauthoryear{}{}]{Choi06}
      Choi P. I., Yan L., Im M. et al., 2006, ApJ, 637, 227 
    
    \bibitem[\protect\citeauthoryear{}{}]{Condon92}
      Condon J. J., 1992, ARA\&A, 30, 575
    
    \bibitem[\protect\citeauthoryear{}{}]{Condon02}
      Condon J. J., Cotton W. D., Broderic J. J., 2002, ApJ, 124, 675
    
    \bibitem[\protect\citeauthoryear{}{}]{Concelice03}
      Conselice C. J., 2003, ApJS, 147, 1
    
    \bibitem[\protect\citeauthoryear{}{}]{Conselice07}
      Conselice C. J., Bundy K., Trujillo I. et al., 2007, MNRAS, 381, 962 (C07)
    
    \bibitem[\protect\citeauthoryear{}{}]{Conselice08a}
      Conselice C. J., Bundy K., Vivian U., Eisenhardt P., Lotz J., Newman J., 2008a, MNRAS, 383, 1366
    
    \bibitem[\protect\citeauthoryear{}{}]{Conselice08b}
      Conselice C. J., Rajgor S., Myers R. 2008b, MNRAS, 386, 909
    
    \bibitem[\protect\citeauthoryear{}{}]{Concelice09}
      Conselice C. J., Yang C., Bluck A. F. L., 2009, MNRAS, 394, 1956
    
    \bibitem[\protect\citeauthoryear{}{}]{Concelice11}
      Conselice C. J., Bluck A. F. L., Buitrago F. et al., 2011, MNRAS, 413, 80
    
    \bibitem[\protect\citeauthoryear{}{}]{Daddi07}
      Daddi E., Dickinson M., Morrison G. et al., 2007a, ApJ, 670, 156 (D07)
    
    \bibitem[\protect\citeauthoryear{}{}]{Daddi07b}
      Daddi E., Alexander D. M., Dickinson M. et al., 2007b, ApJ, 670, 173
    
    \bibitem[\protect\citeauthoryear{}{}]{Davis03}
      Davis M., Faber S. M., Newman J. et al., 2003, SPIE, 4834, 161
    
    \bibitem[\protect\citeauthoryear{}{}]{Davis07}
      Davis M., Guhathakurta P., Konidaris N. P. et al., 2007, ApJ, 660, 1
    
    \bibitem[\protect\citeauthoryear{}{}]{Doherty}
      Doherty M., Bunker A., Sharp R., Dalton G., Parry I., Lewis I., 2006, MNRAS, 370, 331
    
    \bibitem[\protect\citeauthoryear{}{}]{Dong08}
      Dong X., Wang T., Wang J., Yuan W., Zhou H., Dai, H., Zhang, K., 2008, MNRAS, 383, 581
    
    \bibitem[\protect\citeauthoryear{}{}]{Erb2006}
      Erb D. K., Steidel C. C., Shapley A. E., Pettini M., Reddy N. A., Adelberger K. L., 2006, ApJ, 647, 128
    
    \bibitem[\protect\citeauthoryear{}{}]{Fasano87}
      Fasano G., Franceschini A., 1987, MNRAS, 225, 155
    
    \bibitem[\protect\citeauthoryear{}{}]{Fontana04}
      Fontana A., Pozzetti L., Donnarumma I. et al., 2004, A\&A, 424, 23
    
    \bibitem[\protect\citeauthoryear{}{}]{ForsterSchreiber09}
      F\"{o}rster Schreiber N. M., Genzel R., Bouché N. et al., 2009, ApJ, 706, 1364
    
    \bibitem[\protect\citeauthoryear{}{}]{Gallagher89}
      Gallagher J. D., Hunter D. A., Bushouse, H., 1989, AJ, 97, 700
    
    \bibitem[\protect\citeauthoryear{}{}]{Gallego97}
      Gallego J., Zamorano J., Rego M., Vitores A. G., 1997, ApJ, 475, 502
    
    \bibitem[\protect\citeauthoryear{}{}]{Garn10}
      Garn T., Sobral D., Best P. N. et al., 2010, MNRAS, 402, 2017 
    
    \bibitem[\protect\citeauthoryear{}{}]{Giavalisco04}
      Giavalisco M., Lee K.-S., Ferguson H. C. et al., 2004, AAS, 36, 1497
    
    \bibitem[\protect\citeauthoryear{}{}]{Glazebrook99}
      Glazebrook K., Blake C., Economou F., Lilly S., Colless M., 1999, MNRAS, 306, 843
    
    \bibitem[\protect\citeauthoryear{}{}]{Glazebrook04}
      Glazebrook K., Abraham R. G., McCarthy P. J. et al., 2004, Nature, 430, 181
    
    \bibitem[\protect\citeauthoryear{}{}]{Haarsma00}
      Haarsma D. B., Partridge R. B., Windhosrt R. A., Richards E. A., 2000, ApJ, 544, 641
    
    \bibitem[\protect\citeauthoryear{}{}]{Haggard10}
      Haggard D., Green P. J., Anderson S. F., Constantin A., Aldcroft T. L., Kim D.-W., Barkhouse W. A., 2010, ApJ, 723, 1447
    
    \bibitem[\protect\citeauthoryear{}{}]{Hayashi09}
      Hayashi M., Motohara K., Shimasaku K. et al., 2009, ApJ, 691, 140
    
    \bibitem[\protect\citeauthoryear{}{}]{Helou86}
      Helou G., 1986, ApJ, 311, L33
    
    \bibitem[\protect\citeauthoryear{}{}]{Heavens04}
      Heavens A. F., Panter B., Jimenez R., Dunlop J., 2004, Nature, 428, 625
    
    \bibitem[\protect\citeauthoryear{}{}]{Hopkins04}
      Hopkins A. M., 2004, ApJ, 615, 209
    
    \bibitem[\protect\citeauthoryear{}{}]{Hopkins00}
      Hopkins A. M., Connolly A. J., Szalay A. S., 2000, AJ, 120, 2843
    
    \bibitem[\protect\citeauthoryear{}{}]{Hopkins01}
      Hopkins A. M., Connolly A. J., Haarsma D. B., Cram L. E., 2001, AJ, 122, 288
    
    \bibitem[\protect\citeauthoryear{}{}]{Ilbert10}
      Ilbert O., Salvato M., Le Floc'h E. et al., 2010, ApJ, 709, 644
    
    \bibitem[\protect\citeauthoryear{}{}]{Ivison07}
      Ivison R. J., Chapman S. C., Faber S. M. et al., 2007, ApJ 660, L77
    
    \bibitem[\protect\citeauthoryear{}{}]{Ivison10}
      Ivison  R. J., Alexander  D. M., Biggs  A. D. et al., 2010, MNRAS, 402, 245
    
    \bibitem[\protect\citeauthoryear{}{}]{Jansen01}
      Jansen R. A., Marijn F., Fabricant D., 2001, ApJ, 551, 825
    
    \bibitem[\protect\citeauthoryear{}{}]{Johnson06}
      Johnson B. D., Schiminovich D., Seibert M. et al., 2006, ApJ, 644, L109
    
    \bibitem[\protect\citeauthoryear{}{}]{Johnson07}
      Johnson B. D., Schiminovich D., Seibert M. et al., 2007, ApJS, 173, 377
    
    \bibitem[\protect\citeauthoryear{}{}]{Kauffmann03a}
      Kauffmann G., Heckman T. M., White S. D. M. et al., 2003a, MNRAS, 341, 33
    
    \bibitem[\protect\citeauthoryear{}{}]{Kauffmann03b}
      Kauffmann G., Heckman T. M., Tremonti C. et al., 2003b, MNRAS, 346, 1055
    
    \bibitem[\protect\citeauthoryear{}{}]{Kennicutt83}
      Kennicutt R. C., 1983, ApJ, 272, 54,
    
    \bibitem[\protect\citeauthoryear{}{}]{Kennicutt98a}
      Kennicutt R. C., 1998a, ARA\&A, 36, 189
    
    \bibitem[\protect\citeauthoryear{}{}]{Kennicutt98b}
      Kennicutt R. C., 1998b, ApJ, 498, 541
      
    \bibitem[\protect\citeauthoryear{}{}]{Kennicutt09}
      Kennicutt R. C., Hao C.-N., Calzetti D. et al., 2009, ApJ, 703, 1672 (K09)
    
    \bibitem[\protect\citeauthoryear{}{}]{Kewley04}
      Kewley L. J., Geller M. J., Jansen R. A., 2004, AJ, 127, 2002
    
    \bibitem[\protect\citeauthoryear{}{}]{Lacy04}
      Lacy M., Storrie-Lombardi L. J., Sajina  A. et al., 2004, ApJS, 154, L166
    
    \bibitem[\protect\citeauthoryear{}{}]{LonsdalePersson87}
      Lonsdale Persson C. J., Helou G. X., 1987, ApJ, 314, 513
    
    \bibitem[\protect\citeauthoryear{}{}]{Madau98}
      Madau P., Pozzetti L., Dickinson M., 1998, ApJ, 498, 106
    
    \bibitem[\protect\citeauthoryear{}{}]{Maraston06}
      Maraston C., Daddi E., Renzini A., Cimatti A., Dickinson M., Papovich C., Pasquali A., Pirzkal, N., 2006, ApJ, 652, 85
    
    \bibitem[\protect\citeauthoryear{}{}]{McKee97}
      McKee C. F., Williams J. P., 1997, ApJ, 476, 144
    
    \bibitem[\protect\citeauthoryear{}{}]{Noeske07a}
      Noeske K. G., Weiner B. J., Faber S. M. et al., 2007a, ApJ, 660, L43 (N07)
    
    \bibitem[\protect\citeauthoryear{}{}]{Noeske07b}
      Noeske K. G., Faber S. M., Weiner B. J. et al. 2007b, ApJ, 660, L47
    
    \bibitem[\protect\citeauthoryear{}{}]{Osterbrock89}
      Osterbrock  D. E. 1989, Astrophysics of Gaseous Nebulae and Active Galactic Nuclei, University Science Books, Mill Valley, CA
    
    \bibitem[\protect\citeauthoryear{}{}]{Pannella09}
      Pannella  M., Carilli  C. L., Daddi  E. et al., 2009, ApJ, 698 116L
    
    \bibitem[\protect\citeauthoryear{}{}]{Panter07}
      Panter B., Jimenez R., Heavens A. F., Charlot S., 2007, 378, 1550
    
    \bibitem[\protect\citeauthoryear{}{}]{Papovich06}
      Papovich C., Moustakas L. A., Dickinson M. et al., 2006, ApJ, 640, 92
    
    \bibitem[\protect\citeauthoryear{}{}]{Papovich07}
      Papovich C., Rudnick G., Le Floc'h E. et al., 2007, ApJ, 668, 48
    
    \bibitem[\protect\citeauthoryear{}{}]{PerezGonzalez08a}
      P\'{e}rez-Gonz\'{a}lez P. G., Rieke G. H., Villar V. et al., 2008a, ApJ, 675, 234

    \bibitem[\protect\citeauthoryear{}{}]{PerezGonzalez08b}
      P\'{e}rez-Gonz\'{a}lez P. G., Trujillo I., Barro G., Gallego J., Zamorano J., Conselice C. J., 2008b, ApJ, 687, 50

    \bibitem[\protect\citeauthoryear{}{}]{Petrosian72}
      Petrosian V., Silk J., Field G. B., 1972, ApJ, 177, 69
    
    \bibitem[\protect\citeauthoryear{}{}]{RodriguezEugenio}
      Rodr\'{\i}guez-Eugenio N. Noeske K. G., Acosta-Pulido J., Barrena R., Prada F., Manchado A., 2007, IAUS, 235, 417
    
    \bibitem[\protect\citeauthoryear{}{}]{Rosa-Gonzalez02}
      Rosa-Gonz\'{a}lez D., Terlevich E., Terlevich R., 2002, MNRAS, 332, 283
    
    \bibitem[\protect\citeauthoryear{}{}]{Rowan-Robinson89}
      Rowan-Robinson M., Crawford J., 1989, MNRAS, 238, 523
    
    \bibitem[\protect\citeauthoryear{}{}]{Salim07}
      Salim  S., Rich  R. M., Charlot  S. et al., 2007, ApJS, 173, 267
    
    
    \bibitem[\protect\citeauthoryear{}{}]{Schiminovich07}
      Schiminovich D., Wyder T. K., Martin D. C. et al., 2007, ApJS, 173, 315
    
    \bibitem[\protect\citeauthoryear{}{}]{Seibert05}
      Seibert M., Martin D. C., Heckman T. M. et al., 2005, ApJ, 619, L55
    
    \bibitem[\protect\citeauthoryear{}{}]{Silvernab09}
       Silverman J. D., Kova\u{c} K., Knobel C., et al., 2009, ApJ, 695, 171
    
    \bibitem[\protect\citeauthoryear{}{}]{Smail97}
      Smail I., Ivison R. J., Blain A. W., 1997, ApJ, 490, L5
    
    \bibitem[\protect\citeauthoryear{}{}]{Sobral10}
      Sobral D., Best P. N., Smail I., Geach J. E., Cirasuolo M., Garn T., Dalton G. B., 2010, arXiv:1007.2642v1
    
    \bibitem[\protect\citeauthoryear{}{}]{Spitzer78}
      Spitzer L., 1978, Physical Processes in the Interstellar Medium, Wiley, New York
    
    \bibitem[\protect\citeauthoryear{}{}]{Sullivan01}
      Sullivan M., Mobasher B., Chan B., Cram L., Ellis R., Treyer M., Hopkins A., 2001, ApJ, 558, 72
    
    \bibitem[\protect\citeauthoryear{}{}]{Treister05}
      Treister E., Urry C. M., Van Duyne J. et al., 2005, ApJ, 640, 603
    \bibitem[\protect\citeauthoryear{}{}]{Tresse99}
      Tresse L., Maddox S. J., Loveday J., Singleton C., 1999, MNRAS, 310, 262
    
    \bibitem[\protect\citeauthoryear{}{}]{Tresse02}
      Tresse L., Maddox S. J., Le Fèvre O., Cuby J.-G., 2002, MNRAS, 337, 369
    
    \bibitem[\protect\citeauthoryear{}{}]{Veilleux87}
      Veilleux S., Osterbrock D. E., 1987, ApJS, 63, 295
    
    \bibitem[\protect\citeauthoryear{}{}]{Wang06}
      Wang J. X., Jiang P., 2006, ApJ, 646, L103
    
    \bibitem[\protect\citeauthoryear{}{}]{Yan99}
      Yan L., McCarthy P. J., Feudling W., Teplitz H. I., Malumuth E. M., Weymann R. J., Malkam M. A., 1999, ApJ, 519, L47

  \end{thebibliography}
\end{document}